\documentclass{aastex63}

\submitjournal{ApJ}

\shorttitle{Two-fluid Modeling of Acoustic Wave Propagation}
\shortauthors{Zhang et al.}

\begin{document}

\title{Two-fluid Modeling of Acoustic Wave Propagation in Gravitationally Stratified Isothermal Media}

\correspondingauthor{Fan Zhang}
\email{fan.zhang@kuleuven.be, fan.zhang@mail.be}

\author[0000-0002-9425-994X]{Fan Zhang}
\affiliation{Centre for mathematical Plasma-Astrophysics, Department of Mathematics, KU Leuven, Celestijnenlaan 200 B, 3001 Leuven, Belgium}

\author[0000-0002-1743-0651]{Stefaan Poedts}
\affiliation{Centre for mathematical Plasma-Astrophysics, Department of Mathematics, KU Leuven, Celestijnenlaan 200 B, 3001 Leuven, Belgium}
\affiliation{Institute of Physics, University of Maria Curie-Sk{\l}odowska, Pl. M. Curie-Sk{\l}odowskiej 5, 20-031 Lublin, Poland}
\author[0000-0003-4017-215X]{Andrea Lani}
\affiliation{Centre for mathematical Plasma-Astrophysics, Department of Mathematics, KU Leuven, Celestijnenlaan 200 B, 3001 Leuven, Belgium}

\author[0000-0001-9438-9333]{B{\l}a{\.{z}}ej Ku{\'{z}}ma}
\affiliation{Institute of Physics, University of Maria Curie-Sk{\l}odowska, Pl. M. Curie-Sk{\l}odowskiej 5, 20-031 Lublin, Poland}
\affiliation{Centre for mathematical Plasma-Astrophysics, Department of Mathematics, KU Leuven, Celestijnenlaan 200 B, 3001 Leuven, Belgium}
\author[0000-0002-0184-2117]{Kris Murawski}
\affiliation{Institute of Physics, University of Maria Curie-Sk{\l}odowska, Pl. M. Curie-Sk{\l}odowskiej 5, 20-031 Lublin, Poland}

\begin{abstract}
To study acoustic wave propagation and the corresponding energy deposition in partially ionized plasmas, we use a two-fluid computational model which treats neutrals and charged particles (electrons and ions) as two separate fluids. This two-fluid model takes into account the ion-neutral collisions, ionization and recombination, allowing us to investigate both the collisional and reactive interactions between uncoupled ions and neutrals in the plasmas. In the present numerical simulations, the initial density is specified to reach hydrostatic equilibrium, and as a comparison, chemical equilibrium is also taken into account to provide a density profile that differs from typical hydrostatic equilibrium profiles. External velocity drivers are then imposed to generate monochromatic acoustic waves. As it is well known, the upwards propagating acoustic waves steepen in gravitationally stratified plasmas due to the exponentially decreasing density, and they heat the plasmas in the nonlinear regimes where kinetic energy is dissipated by shock waves and collisional interactions. In particular, the lower ionization fraction resulting from the present initial chemical equilibrium significantly enhances the heating efficiency. Moreover, the ionization process absorbs a significant amount of energy, and the decoupling between ions and neutrals is also enhanced while considering ionization and recombination. Therefore, simulations without considering ionization and recombination may overestimate the overall heating effects but also underestimate the energy dissipation. The results also suggest that a more accurate ionization and recombination model could be essential for improving the modeling of partially ionized plasmas. 
\end{abstract}


\section{Introduction} \label{sec:intro}

The solar chromosphere is a thin layer (about $2,000\;$km thick) of the solar atmosphere which is situated above the photosphere and below the transition region, and this layer is full of complex and interesting phenomena. For instance, the temperature profile of the chromosphere is counter-intuitive. Normally, it is straightforward to expect that the temperature decreases as the height increases, as suggested by the second law of thermodynamics and presuming all thermal energy to be produced in the solar interior. However, in reality, the temperature of the chromosphere increases from a minimum of around $4,200\;$K, at the region right above the photosphere, to a maximum of around 
$30,000\;$K, at the edge with the transition region \citep{Vernazza1981}. In fact, the chromosphere not only has a temperature profile that increases with distance from the Sun, it also loses a significant amount of energy through strong radiation (stronger than that of the corona), which means that extra energy compensation is required {\citep{Athay1976,Withbroe1977}}.
Therefore, revealing the fundamental energy transportation mechanism(s)  in the chromosphere is a long-standing topic of solar physics  {and has been discussed in numerous articles, such as \citet{Biermann1946,Schwarzschild1948,Ulmschneider1971,Ulmschneider1971_2,Ulmschneider1977,Kalkofen1977,Ulmschneider1977_2,Schmieder1979,Fossum2005,Ulmschneider2005,Maneva_2017,Yalim2020}, to name a few}.

It is actually challenging to numerically investigate the solar chromosphere, partly because the lower solar atmosphere is only {partially} ionized, which means that neutrals play an important role in wave propagation \citep{Krasnoselskikh2010,Soler2015,Khomenko2016}, leading to a necessity of multi-fluid modeling since typical ideal magnetohydrodynamics (MHD) or single-fluid approximation is not sufficient to describe the complex physics \citep{Zaqarashvili2011,Zaqarashvili2012,Khomenko2012,Soler2013a,Soler2013,Khomenko2014,Khomenko2016,Alharbi2021}. Moreover, as one of the essential interaction mechanisms between charged particles (electrons and ions) and neutrals, {the contribution of collisional interactions is considered as  an important factor while modeling} MHD wave propagation, damping and heating in partially ionized plasmas  \citep{Khomenko2012,Soler2013,Khomenko2016,Martinez-Gomez2017,Martinez-Gomez2018,PopescuBraileanu2019}, and the influence may be even more important for waves in nonlinear regimes \citep{PopescuBraileanu2019}.
 
In general, because of their higher accuracy in the description of partially ionized plasmas, multi-fluid models are used to numerically model various physical processes, especially including the chromospheric magnetic reconnection \citep{Leake2012,Leake2013,Murphy2015,Laguna2017} and one-dimensional \citep{Martinez-Gomez2017,Martinez-Gomez2018,Wojcik2018,Kuzma2019,PopescuBraileanu2019} or multi-dimensional wave propagation \citep{Soler2017,Kuzma2017,Maneva_2017,Soler2019a,Wojcik2019a,Wojcik2020,PopescuBraileanu2021} in the solar chromosphere or solar prominences. In particular, we are interested in the multi-fluid numerical modeling of wave propagation in the lower solar atmosphere, including the photosphere and the chromosphere, since wave propagation is deemed to be an important mechanism that contributes to the heating of the lower solar atmosphere. Moreover, the photosphere is a dynamical thin layer {(about $500\;$km thick)} that hosts a wide range of oscillations, and thus it becomes a source of waves propagating into the upper atmosphere including the neighbouring solar chromosphere. Therefore, the wave propagation/damping mechanism in these regions is particularly interesting for investigating the heating process of the chromosphere, and also the upper atmospheric layers.

To proceed the numerical modeling of wave propagation in the solar atmosphere, {appropriate initial/background fields are required. In reality, as mentioned above, the photosphere and the chromosphere are both highly dynamical, and thus a granulation associated field may be used to initiate the simulations \citep{Khomenko2018,Morales2020}. Whereas, in order to specifically investigate the wave propagation mechanism in an ideal background field,  a simplified initial hydrostatic equilibrium model may be  considered, frequently} corresponding to the one-dimensional average gravitationally stratified quiet-solar atmosphere \citep{Vernazza1981}. This results in a gravitationally stratified medium of which the density profile decreases exponentially when moving upward, which causes the waves generated in the {weakly ionized} photosphere to steepen in the partially ionized chromosphere. Therefore, there are abundant physical phenomena preferably to be described by  multi-fluid models. For instance, by using a three-fluid model, \citet{Soler2017} modelled torsional Alfv\'{e}n waves which are driven below the
photosphere and propagate up to the corona, revealing the reflection, transmission and damping mechanisms which depend on wave frequencies, and estimating the chromospheric heating rates.  \citet{Maneva_2017} first investigated magneto-acoustic wave propagation in the solar chromosphere while considering the effects of impact ionization and radiative recombination. \citet{PopescuBraileanu2019} provided a detailed explanation for the propagation and damping of fast magneto-acoustic waves and shocks in the solar chromosphere, and the results suggested that the decoupling between charged particles and neutrals leads to collisional wave damping and eventually causes an increase of the plasma temperature. Moreover, acoustic wave propagation without magnetic effects were also investigated by multi-fluid modeling \citep{Wojcik2018,Kuzma2019}.

{Obviously, among these wave propagation processes which may be important for the chromospheric heating problem, acoustic wave propagation in a one-dimensional gravitationally stratified plasma is a highly idealised and simplified model for investigating the heating of the (non-magnetic) solar atmosphere. Whereas, it also provides a scenario for clearly showing the effects of different physical mechanisms, and in fact, acoustic wave propagation is considered to be an important process of energy transport in the lower solar atmosphere. \citet{Biermann1946} and \citet{Schwarzschild1948} first suggested that the acoustic waves generated by the granulation may transport mechanical energy to heat the solar chromosphere.  
\citet{Ulmschneider1970,Ulmschneider1971,Ulmschneider1971_2} discussed acoustic waves generated in the convection zone, and concluded that the acoustic waves and the resulting shock waves in the chromosphere provide mechanical heating that compensates the net chromospheric radiation loss. \citet{Stein1972,Stein1973} further investigated acoustic pulse and periodic wave train, while taking into account the effects of ionization by using the Saha's equation \citep{D.Sc.1920}.
Later a systematic research had been done by \citet{Ulmschneider1977,Kalkofen1977,Ulmschneider1977_2}, who investigated acoustic waves in the solar atmosphere, and supported the short period acoustic heating theory of the chromosphere. Of course, these one-dimensional non-magnetic analyses do not necessarily represent the realistic chromosphere, since magnetic field and/or multi-dimensional effects need to be taken into account \citep{Fossum2005,Ulmschneider2005}. Therefore, as a conclusive explanation about the acoustic wave heating is not yet provided \citep{Kalkofen2007}, more accurate and realistic numerical models including radiation \citep{Bard2010}, multi-dimensional effects \citep{Kalkofen2010} and/or multi-fluid effects \citep{Kuzma2019}, are still required. }

In this work, we further investigate {one-dimensional} acoustic wave propagation  by using a two-fluid plasma-neutral model to provide new insights of the wave damping mechanism and the heating process {in partially ionized plasmas}. In fact, while using multi-fluid modeling to investigate wave propagation,  only a few attempts have been made to take into account the effects of the ionization and recombination processes \citep{Reep2016,Maneva_2017,Snow2021}. Yet, the partially ionized plasma in the chromosphere does not fulfil local thermodynamic equilibrium (LTE), resulting in different temperature profiles of charged particles and neutrals, and due to ionization and recombination, the plasma may vary from weakly ionized to fully ionized while the temperature is increasing, leading to important effects on the properties of the plasma. For instance, the falling-off ionization fraction may enhance the effectiveness of ion-neutral friction \citep{Reep2016}. Therefore, in this work,  following the study of \citet{Maneva_2017}, who further exploited the two-fluid model developed by \citet{Leake2012}, we numerically investigate acoustic wave propagation in {isothermal plasmas, which have similar quantities as in the chromosphere but are simplified as pure hydrogen}, while excluding the charge exchange terms \citep{Meier2012} in the governing equations, but including the elastic collisions \citep{Vranjes2013} and the chemical reactions (impact ionization and radiative recombination) \citep{Leake2012}.
Moreover, we explicitly impose initial hydrostatic equilibrium and/or chemical equilibrium, and thus wave propagation and the corresponding heating process can be investigated without initial hydrodynamic or chemical imbalances. The remainder of this paper is organized as follows. In section~\ref{sec:style}, the two-fluid numerical model equations and the boundary and initial conditions are introduced in detail. In section~\ref{sec:results}, the numerical results are provided and explained. Some concluding remarks are presented in the final section.

\section{Numerical Models and Computational Setup} \label{sec:style}

As mentioned above, in the present study we use a two-fluid numerical model {to investigate the acoustic wave propagation, which suppose to start in the photosphere. However, it must be noted that this largely simplified model assuming a pure hydrogen plasma is not suitable for modeling the photosphere where heavy elements are the major source of electrons, but should be used for the upper chromoshpere or even prominences. For instance, a similar two-fluid model is used by \citet{PopescuBraileanu2019} for numerically investigating the chromosphere. Therefore, the present model is to be further improved later on for accurately modeling the lower solar atmosphere. }

In the present two-fluid model, the ions and electrons together are described by using a single-fluid simplification, assuming that they have the same temperature and velocity, and neglecting the electron mass effects. The second fluid has independent temperature and velocity, modeling the behaviour of neutrals. Elastic collisions, impact ionization, and radiative recombination are considered in modeling the interaction mechanisms between these two fluids \citep{Braginskii1965}. The governing equations of this model, the numerical schemes, the boundary and initial conditions are introduced in the following subsections. Moreover, we simply focus on acoustic waves, without considering magnetic field, and thus the model equations can be further simplified.

\subsection{Governing Equations}

In this work, the two-fluid Euler equations modeling the partially ionized plasma, and including the continuity, momentum, and
energy equations for the ions and the neutrals can be formally written as
\begin{eqnarray}
&\frac{\partial \varrho_{\text{s}}}{\partial t}+\nabla\cdot(\varrho_{\text{s}}\mathbf{v}_{\text{s}})=m_{\text{s}}{S}_{\text{s}}, \nonumber \\
&\frac{\partial (\varrho_{\text{s}}\mathbf{v}_{\text{s}})}{\partial  t}+\nabla\cdot(\varrho_{\text{s}}\mathbf{v}_{\text{s}}\mathbf{v}_{\text{s}}+p_{\text{s}})=\varrho_{\text{s}}\mathbf{g}+\mathbf{R}_{\text{s}}, \nonumber \\
&\frac{\partial e_{\text{s}}}{\partial t}+\nabla\cdot\left(e_{\text{s}}\mathbf{v}_{\text{s}}+p_{\text{s}}\mathbf{v}_{\text{s}}\right) =\varrho_{\text{s}}\mathbf{v}_{\text{s}}\cdot\mathbf{g}+{M}_{\text{s}},
\end{eqnarray}
\noindent where the subscript $\text{s}=\{\text{i,}\,\text{n}\}$, indicates the species (\textbf{i}ons or \textbf{n}eutrals), $t$ is time, $\varrho$ denotes the mass density, $\mathbf{v}$ is the velocity, $e$ is the total energy including kinetic energy $\frac{1}{2}\varrho v^2$ and internal energy $\frac{p}{\gamma-1}$ with the ratio of specific heats $\gamma=\frac{5}{3}$, $p$ is the pressure, $m$ is the molecular mass, and the constant gravitational acceleration is specified as $\mathbf{g}=(0,g,0)=(0,-274.78\;$ms$^{-2},0)$. An ideal gas equation of state, i.e.\ $p=nk_{\text{B}}T$, is used to close the set of equations, where $T$ is the temperature, $n$ is the number density, and $k_{\text{B}}$ is the Boltzmann constant. Since the magnetic field is not taken into account in this paper, the governing equations are formally the same for both ions and neutrals, and the only term to be further clarified is the ion pressure  ($p_{\text{i}}$), which  actually includes the electron pressure ($p_{\text{e}}$), resulting from the single-fluid ion+electron description.

The interactions between ions and neutrals are described by the source terms.
The first source term, ${S}_{\text{s}}$, includes the impact ionization and radiative recombination, i.e.\
\begin{eqnarray}
{S}_{\text{i}}=-{S}_{\text{n}}=\Gamma^{\text{ion}}-\Gamma^{\text{rec}}.
\end{eqnarray}
\noindent The ionization and recombination coefficients, $\Gamma^{\text{ion}}$ and $\Gamma^{\text{rec}}$,  follow the expressions \citep{Cox1969}
\begin{eqnarray} \label{eq:ChemEquilibrium}
&\Gamma^{\text{ion}}=n_{\text{i}}n_{\text{n}}I, \nonumber\\
&\Gamma^{\text{rec}}=n_{\text{i}}^2R,
\end{eqnarray}
\noindent where $n_{\text{s}}$ corresponds to the number density of \textbf{i}ons or \textbf{n}eutrals, and
\begin{eqnarray}
&I=2.34\cdot 10^{-14}\left(\beta^{-\frac{1}{2}}\exp(-\beta)\right)\,\text{m}^3 \text{s}^{-1}, \nonumber\\
&R=5.20\cdot 10^{-20}\sqrt{\beta}\left(0.4288+0.5\ln(\beta)+0.4698\beta^{-1/3}\right)\,\text{m}^3 \text{s}^{-1}.
\end{eqnarray}
\noindent More specifically, $\beta=A\cdot \Phi_{\text{ion}}/T_e$ is a non-dimensional function that describes the temperature dependence of the ionization and recombination processes, $T_e$ is the electron temperature in eV and, in fact, equals the ion temperature in the present model, $\Phi_{\text{ion}}=13.6\,$eV is the ionization energy of hydrogen, and the constant $A=0.6$ takes into account the contribution of heavy ions in the present pure hydrogen plasma \citep{Maneva_2017}.

The second source term, $\mathbf{R}_{\text{s}}$, is the momentum source term which includes the contributions of both the collisions and the ionization and recombination. Therefore, it is written as
\begin{eqnarray}
\mathbf{R}_{\text{i}}=-\mathbf{R}_{\text{n}}=&\mathbf{R}_{\text{in}}+\Gamma^{\text{ion}}m_{\text{i}}\mathbf{v}_{\text{n}}-\Gamma^{\text{rec}}m_{\text{i}}\mathbf{v}_{\text{i}}, \end{eqnarray}
\noindent where the first term in the right-hand side describes the collisional momentum exchange, and the second and third terms are the momentum source terms resulting from the ionization and recombination, respectively.  More specifically, the elastic collisions between ions and neutrals are described by \citep{Leake2012,Leake2013}
\begin{eqnarray}
\mathbf{R}_{\text{in}}=m_{\text{in}}n_{\text{i}}n_{\text{n}}\Sigma_{\text{in}}\sqrt{\frac{8k_{\text{B}}T_{\text{in}}}{\pi m_{\text{in}}}}\left(\mathbf{v}_{\text{n}}-\mathbf{v}_{\text{i}}\right),
\end{eqnarray}
\noindent or
\begin{eqnarray}
\mathbf{R}_{\text{in}}=m_{\text{in}}n_{\text{i}}\nu_{\text{in}}\left(\mathbf{v}_{\text{n}}-\mathbf{v}_{\text{i}}\right),
\end{eqnarray}
\noindent where $\nu_{\text{in}}=n_{\text{n}}\Sigma_{\text{in}}\sqrt{\frac{8k_{\text{B}}T_{\text{in}}}{\pi m_{\text{in}}}}$ is the ion-neutral collision frequency, $m_{\text{in}}=\frac{m_{\text{i}}m_{\text{n}}}{m_{\text{i}}+m_{\text{n}}}$ is the reduced mass, $T_{\text{in}}=(T_{\text{i}}+T_{\text{n}})/{2}$ is the average temperature of ions and neutrals, and $\Sigma_{\text{in}}=1.16\times 10^{-18}\,\text{m}^2$ is the collisional cross section taken from \citet{Leake2013}.

The third source term, ${M}_{\text{s}}$, describes the energy exchange and production due to collisions, ionization and recombination, and is written as
\begin{eqnarray}
{M}_{\text{i}}=-{M}_{\text{n}}={M}_{\text{in}}={Q}_{\text{in}}+{Q}^{\text{ion}}-{Q}^{\text{rec}}+\Gamma^{\text{ion}}\frac{1}{2}m_{\text{i}}v_{\text{n}}^2-
 \Gamma^{\text{rec}}\frac{1}{2}m_{\text{i}}v_{\text{i}}^2,
\end{eqnarray}
\noindent where the {collisional energy source term is}  defined as
\begin{eqnarray}  \label{eq:collisional}
{Q}_{\text{in}}=\frac{1}{2}m_{\text{in}}n_{\text{i}}\nu_{\text{in}} ({v}^2_{\text{n}}-{v}^2_{\text{i}})+3\frac{m_{\text{in}}}{m_{\text{H}}}n_{\text{i}}\nu_{\text{in}}k_{\text{B}}(T_{\text{n}}-T_{\text{i}}),
\end{eqnarray}
\noindent and the thermal energy exchange terms due to the chemical reactions are defined as
 \begin{eqnarray}
&{Q}^{\text{ion}}=\frac{3}{2}\Gamma^{\text{ion}}k_{\text{B}}T_{\text{n}}, \nonumber \\
&{Q}^{\text{rec}}=\frac{3}{2}\Gamma^{\text{rec}}k_{\text{B}}T_{\text{i}}.
\end{eqnarray}

In this paper, we neglect viscosity and radiation, but it should be noted that viscosity is also an important damping mechanism in the dense photosphere and the low chromosphere, and the radiation is {not negligible} for balancing the heating mechanism and resulting in a realistic equilibrium. The present simplification is adopted to specifically investigate the collisional and reactive effects.

\subsection{Numerical Schemes}

The present two-fluid numerical simulations are performed using
COOLFluiD\footnote{\url{https://github.com/andrealani/COOLFluiD/wiki}}, which is an open-source component-based software framework for high-performance scientific and engineering computation \citep{Lani2005,Lani2006,Lani2013}. COOLFluiD integrates state-of-the-art computational models and numerical solvers, including ideal MHD model for space weather prediction \citep{Yalim2011,Lani2014}, a multi-fluid model \citep{Laguna2016} for magnetic reconnection \citep{Laguna2017} and wave propagation \citep{Maneva_2017} in the solar chromosphere, etc. In this work, the fully implicit two-fluid (plasma-neutral) \& Maxwell cell-centered finite volume solver developed by \citet{Laguna2016} within COOLFluiD and applied by \citet{Laguna2017,Maneva_2017}, is further exploited for modeling acoustic wave propagation in {partially ionized} plasmas. Of course, the Maxwell solver is not used here because magnetic field is neglected.

More specifically, the fully implicit temporal solution of the two-fluid equations with stiff source terms (collision, ionization and recombination) is based on the three-point backward Euler scheme which is second-order accurate in time.
The computation of
the convective fluxes of the two-fluid equations is based on the Lax-Friedrichs scheme. The spatial derivatives of primitive variables, including density ($\varrho$), velocity ($\mathbf{v}$) and temperature ($T$), are calculated using the Weighted Least SQuares (WLSQ) linear reconstruction for unstructured grids, resulting in a second-order accurate spatial discretization, and the Barth-Jespersen limiter \citep{Barth1989} is applied with a strict limiting criterion \citep{ZhangThesis2017,Zhang2018} to guarantee the robustness of shock-capturing computation in the present gravitationally stratified media where smooth acoustic waves may steepen to shocks. The resulting sparse algebraic system is solved
by using the Generalized Minimal RESidual (GMRES) method \citep{Saad1986} with the restricted additive Schwarz preconditioner \citep{Widlund1987}, and the methods are provided by the Portable, Extensible Toolkit for Scientific computation (PETSc)\footnote{\url{https://www.mcs.anl.gov/petsc/}}. 
 
\subsection{Boundary Conditions and Initial Conditions} \label{sec:BC_IC}

In this work, the quasi one-dimensional numerical box {first covers a region that resembles the spatial scale of the photosphere and chromosphere, namely for $0$ km $ \le y \le 2000\;$km. In this region, 4000 identical cells are used in the vertical direction, which means that each cell covers $500\;$m.} A finer mesh was used and this didn't change the general results, while a coarser mesh could introduce excessive numerical dissipation which changes the pattern of kinetic energy decay to be shown, and thus the present mesh resolution is deemed to be sufficient for resolving the waves being investigated. Above {this region}, namely for
$2000\;\text{km}<y<30,000\;$km,  we gradually stretch the grid, and thus the cell size increases with altitude to absorb or smooth out all the numerical artefacts (waves), which is an effective strategy being used in various numerical simulations of the solar atmosphere \citep{Kuzma2019,Wojcik2020}.

We implement fixed boundary conditions at the bottom of the simulation box, which means that we set all plasma quantities (density and temperature) to their equilibrium values. In the meantime, the velocity driver defining the velocity components at the bottom boundary is imposed according to the following expressions:
\begin{eqnarray}
& {v_{x,\text{s}}}(t)=0,\nonumber \\
 & {v_{y,\text{s}}}(t)=v_0\sin(\frac{2\pi t}{P}),
\end{eqnarray}

\noindent where the subscript $s$ again indicates two species (ions or neutrals) involved in our two-fluid model,  $v_0=100\;$m~s$^{-1}$ is the constant amplitude of the drivers,  {which is assumed to be a reasonable guess of the material speed produced by the motion of granules \citep{Schwarzschild1948}}, and $P$ is the period of the driver, {which is given as $10\;$s, $20\;$s, and $30\;$s, resulting in periodic wave trains of acoustic waves and shocks. In particular, the wave periods are chosen because short period waves were suggested to be responsible for the heating of the chromosphere \citep{Ulmschneider1977_2}, and currently the observational instruments may not be able to fully detect the high-frequency waves \citep{Sobotka2016}, leading to a necessity of using novel numerical models.} At the bottom boundary, the ions and neutrals have the same velocity.
At the top boundary, all the quantities at ghost cells are extrapolated from the internal boundary cells, but the simulations will be stopped before  {the waves reach the top boundary  ($y=30,000\;$km) of the extended buffer zone, where the waves are largely damped by the numerical dissipation}.

All the initial quantities are in hydrostatic equilibrium in {the  region that we are interested in ($0$ km $ \le y \le 2000\;$km)}. In particular, all the initial velocity components are zero, and the other quantities (density and temperature) are given by the functions below:
\begin{eqnarray} \label{eq:Hydro}
&\varrho_{\text{s}}(y)=\varrho_{0,\text{s}} \exp\left(\frac{m_{\text{H}} gy }{ C_{\text{s}} k_{\text{B}} T_{\text{s}}(y)}\right),  \nonumber \\
&T_{\text{s}}(y)=6430\,\text{K},
\end{eqnarray}
\noindent where $\varrho_{0,\text{n}}$ of the following neutral density is chosen to be the value at the bottom of the photosphere from the VAL~C model \citep{Vernazza1981}, and $T$ approximately represents the temperature in the {midchromosphere}. We don't need to further specify the pressure profiles, which are automatically calculated according to the equation of state, {because in our solver, temperature instead of pressure  is used as one of the primitive variables}. It should be noted that $C_{\text{i}}=2$ and $C_{\text{n}}=1$ are, respectively, used for reaching hydrostatic equilibrium for ion and neutral profiles. In this scenario, obviously, the initial chemical equilibrium is violated. 

In order to reach hydrostatic equilibrium and chemical equilibrium at the same time, another density profile is provided by replacing $C_{\text{s}}$ in Eq.~(\ref{eq:Hydro}) by
\begin{eqnarray} \label{eq:Hydro_Chem}
C_{\text{in}}=\frac{n_{\text{i}} C_{\text{i}}+n_{\text{n}} C_{\text{n}}}{n_{\text{i}}+n_{\text{n}}},
\end{eqnarray}
\noindent where the ion and neutral number densities, $n_{\text{i}}$ and $n_{\text{n}}$, are calculated according to Eq.~(\ref{eq:ChemEquilibrium}), by assuming $\Gamma^{\text{ion}}=\Gamma^{\text{rec}}$.  This parameter ($C_{\text{in}}$) is used for calculating both the ion and neutral profiles, thus guaranteeing initial chemical equilibrium. Here, ions and neutrals are respectively not in hydrostatic equilibrium. It is assumed that ions and neutrals are strongly coupled, which is valid in the low solar atmosphere ($y<2000\;$km). Therefore, by using Eq.~(\ref{eq:Hydro_Chem}) in Eq.~(\ref{eq:Hydro}), the imbalanced forces, respectively acting on ions and neutrals, may balance each other, resulting in a quasi-hydrostatic equilibrium, and eventually attaining a hydrostatic+chemical equilibrium initial field.

\begin{figure*}
\gridline{\fig{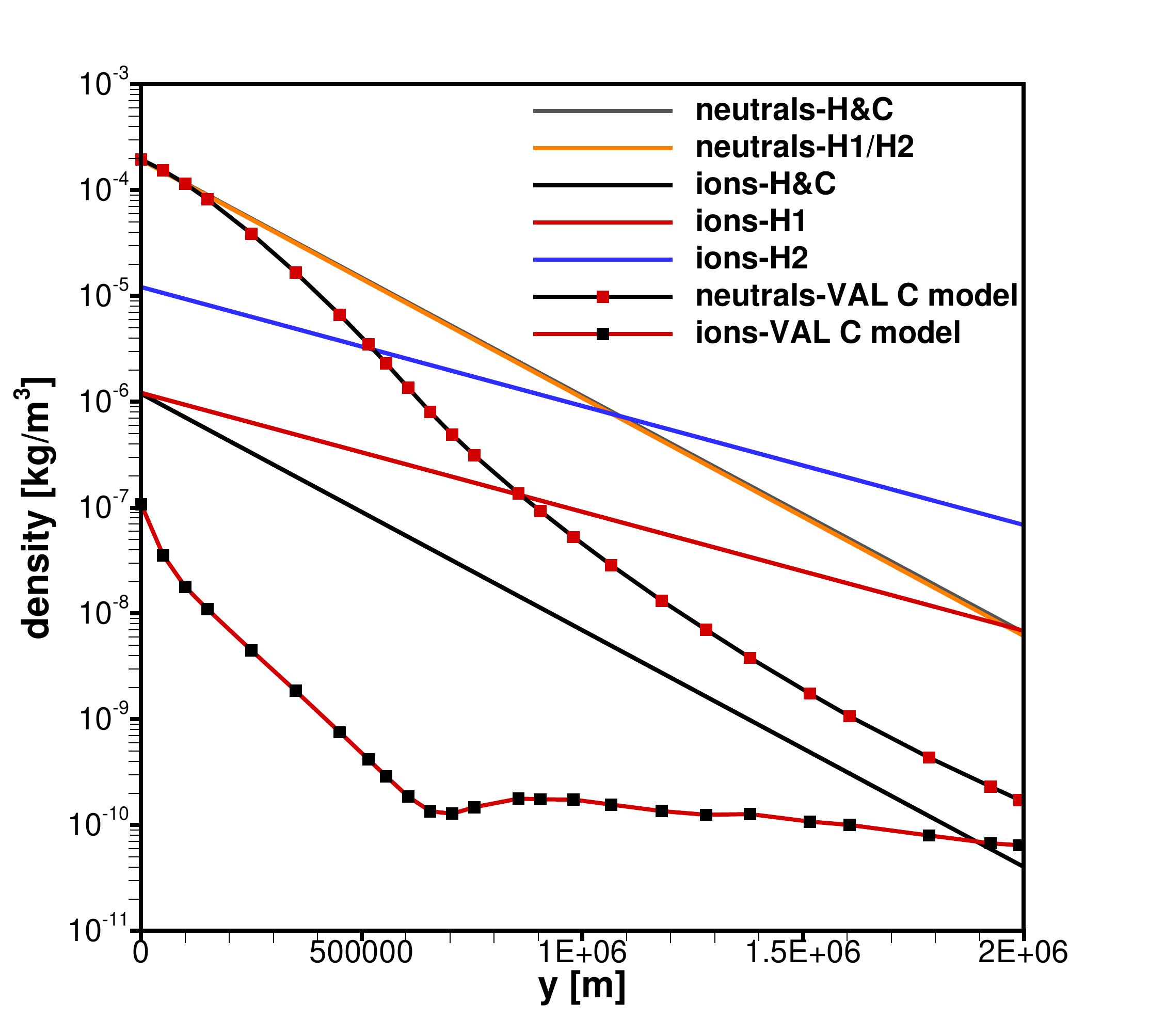}{0.5\textwidth}{(a)}
          \fig{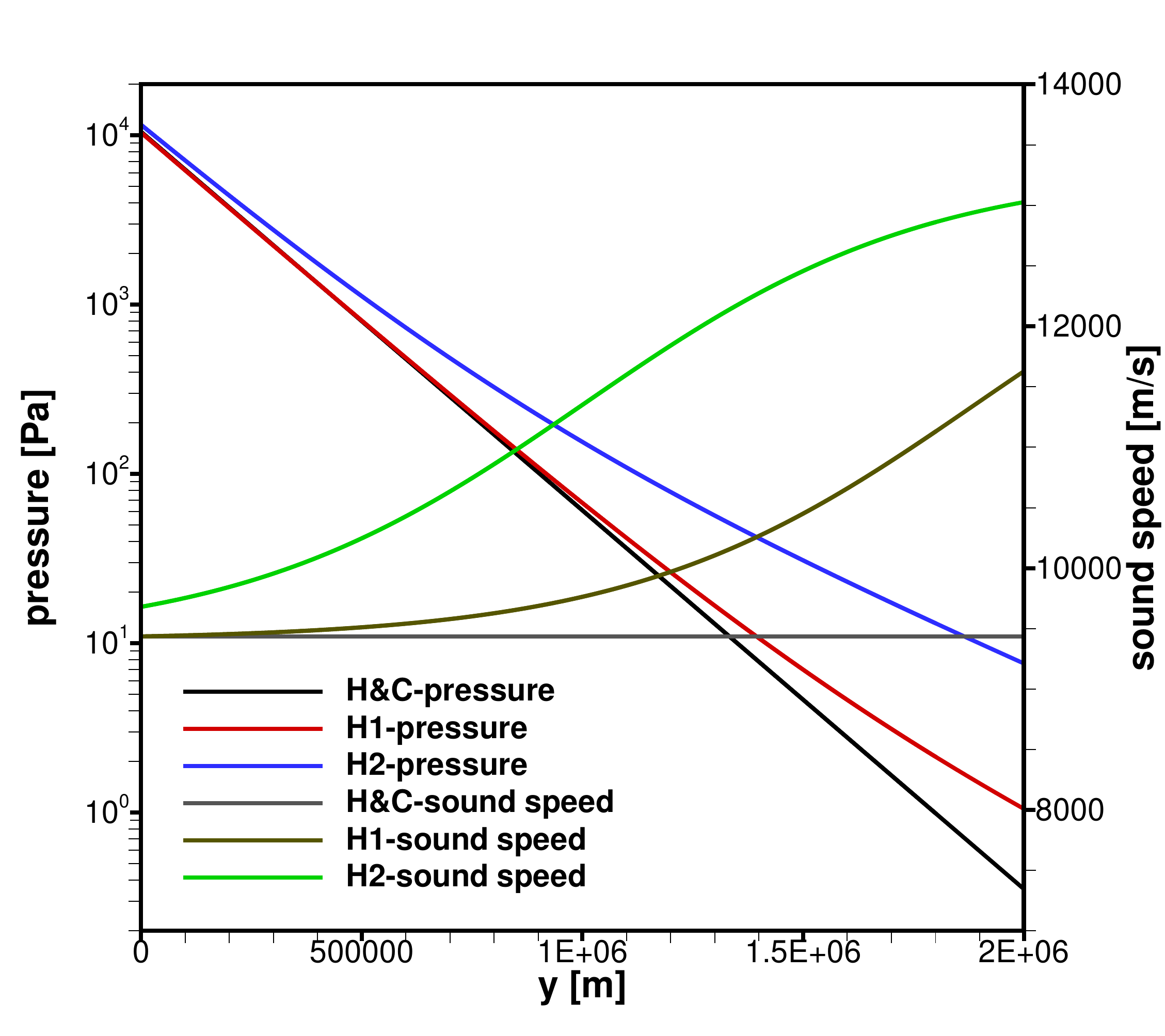}{0.5\textwidth}{(b)}
          } 
\caption{{Initial density, (ion+neutral) pressure and sound speed profiles} satisfying hydrostatic equilibrium and/or chemical equilibrium. H\&C-profile: hydrostatic+chemical equilibrium profile; H1-profile: hydrostatic equilibrium profile; H2-profile: hydrostatic equilibrium profile with higher ion density. {Here,} the density profile of the VAL C model {is calculated while assuming a pure hydrogen plasma}.
\label{fig:density}}
\end{figure*}

In this work, three different density profiles are provided, as shown in Fig.~\ref{fig:density}. In this figure, the two neutral density profiles respectively belong to the hydrostatic equilibrium profiles (H1-profile and H2-profile) and the hydrostatic+chemical equilibrium profile (H\&C-profile). The two different neutral density profiles are actually close because $C_{\text{in}}$  (only) slightly differs from $C_{\text{n}}$. Whereas, a significant difference can be found in the ion density distributions. The $\varrho_{0,\text{i}}$ of the H\&C-profile or the H1-profile is calculated based on the chemical equilibrium assumption, and the $\varrho_{0,\text{i}}$ of the H2-profile is one order magnitude larger than the previous one.
The ion density of the H\&C-profile is decreasing faster, because $C_{\text{in}}$ is significantly smaller than $C_{\text{i}}$, and the H2-profile adopts higher ion density compared with the H1-profile, although these two hydrostatic equilibrium profiles have the same neutral density profile.

Corresponding pressure and sound speed profiles  are also shown. {In particular, the pressure is calculated according to the equation of state, and the sound speed of the partially ionized plasma is given as}
\begin{eqnarray} \label{eq:soundspeed}
c_{\text{in}}(y)=\sqrt{\gamma \frac{p_{\text{i}}(y)+p_{\text{n}}(y)}{\varrho_{\text{i}}(y)+\varrho_{\text{n}}(y)}}.
\end{eqnarray}
\noindent Two hydrostatic equilibrium profiles show higher pressure at higher altitudes, because of the higher ion density. Similarly, because the sound speed of ions(+electrons) is higher than that of neutrals, the sound speed of the partially ionized plasmas becomes higher while the ionization fraction increases. Apparently, the sound speed of the hydrostatic+chemical equilibrium plasma is constant as its ionization fraction is constant. Moreover, the density profile of the VAL C model \citep{Vernazza1981} is also shown. It should be noted {again that, in this work, while calculating all these density profiles, we only consider pure hydrogen plasmas.} We can see that all the present profiles have higher density compared with the VAL C profile. Correspondingly, the pressure and the sound speed also differ from those of the VAL C model. Therefore, the present idealised equilibrium models may only be used for investigating basic wave heating processes in the plasmas, and more realistic models are required for accurately modeling the chromospheric heating problem.

{As in this work we do not try to fully recover the VAL C model, Fig.~\ref{fig:density} still shows an important information frequently being ignored}, which is that the chemical equilibrium itself imposes an extra constraint on the ionization fraction that is obviously important for the interactions between ions and neutrals. As soon as the temperature profile is provided, the ionization fraction is unique in case of the chemical equilibrium state. Whereas, without assuming the initial chemical equilibrium, it is possible to set up an arbitrary amount of hydrostatic equilibrium density profiles and ionization fractions, {resulting in different numerical results}.

Of course, it should be {also} noted that assuming the initial chemical equilibrium may still not represent the realistic solar chromosphere, as it is reported that  the {timescales of} ionization and recombination could be long compared with the hydrodynamic {timescales} \citep{Carlsson1992,Carlsson2002}. Therefore, a non-equilibrium ionization {(NEI)} model \citep{Leenaarts2007} was applied in numerical investigations of the chromosphere \citep{NobregaSiverio2020,Martinez-Sykora2020}, which used a radiative MHD code developed by \citet{Gudiksen2011}. {They have} found that {the simulations assuming LTE and thus excluding the NEI effects
may misestimate the ionization fraction and hence the influence of ambipolar diffusion.} 
In this work, the two-fluid model allows non-LTE effects \citep{Leake2012}, but the ionization/recombination model used here \citep{Cox1969} {still} assumes ionization equilibrium and needs to be further developed {later on for taking into account the NEI effects}.

\section{Numerical Results and Discussion} \label{sec:results}

As introduced in the previous section, we solve the two-fluid system of equations using the fully implicit solver developed by \citet{Laguna2016}, and thus, {theoretically, we are able to assign  the computational time step without any  stability limitation. However, as the sparse linear system resulting from the stiff  two-fluid equations is solved by an iterative method, using a very large time step may deteriorate the convergence, especially while shocks are formed. Moreover, there is a physical constraint on the time step following from the wave periods we are introducing. Therefore,} in our simulations, the time step is gradually increasing from $0.001\;$s to $0.02\;$s in the first 500 time steps, and then we use a constant time step of $0.02\;$s throughout the simulations. This is sufficient as the shortest wave periods considered here are $10\;$s, i.e.\ very well resolved with 500 time steps per period, {ensuring both the accuracy and the convergence of the temporal solutions}.

The effects of taking into account the initial chemical equilibrium are investigated in the first subsection, and the influence of the  ionization and recombination source terms is also discussed. In the second subsection, different velocity  drivers are imposed to further investigate the corresponding energy damping mechanism. It should be noted that, {the driven acoustic waves first reach the upper end of the interested region ($y = 2\;$Mm) at around $200\;$s, and} the results at around $1200\;$s are {illustrated. The waves propagating into the extended buffer zone ($y > 2000\;$km) are largely damped and not reflected, and then quasi-stationary wave structures can be built,} minimising transient disturbances and also reaching sufficient changes (heating) of the plasmas.

\subsection{Initial Chemical Equilibrium and Chemical Reactions} \label{sec:firstPart}

As mentioned previously, most of the two-fluid numerical simulations of the solar chromosphere are conducted {based on the typical hydrostatic equilibrium initial field}, without taking the initial chemical equilibrium {and reactive interactions} into account \citep{Wojcik2018,Kuzma2019,PopescuBraileanu2019,Wojcik2020}, which means that as soon as the ionization and recombination are involved, the chemical non-equilibrium initial field will evolve due to ionization and/or recombination, resulting in disturbances including extra heating or cooling which is not expected while specifically {investigating the wave heating mechanism of the solar atmosphere}. Moreover, while a heating mechanism is imposed, in reality the ionization fraction should be changed correspondingly, {which also results in a change in the collisional interactions}.

Therefore, it is necessary to investigate the influence of including the ionization and recombination. The first effect is that, since the collision term strongly depends on the ion and neutral densities, using the H\&C-profile could already affect the solutions even before including the ionization and recombination processes. In this subsection, we present the numerical results of four different settings. Specifically, collisional simulations without the reactive source terms (ionization and recombination) are performed using all the density profiles, including the H\&C-profile, and resulting in three different numerical results. The only simulation involving reactive source terms uses the H\&C-profile for initialization, avoiding initial chemical imbalance. The velocity driver with $P=10\;$s is used in this subsection for all the simulations.

\begin{figure}[ht!]
 \centering
 \includegraphics[width=0.7\textwidth]{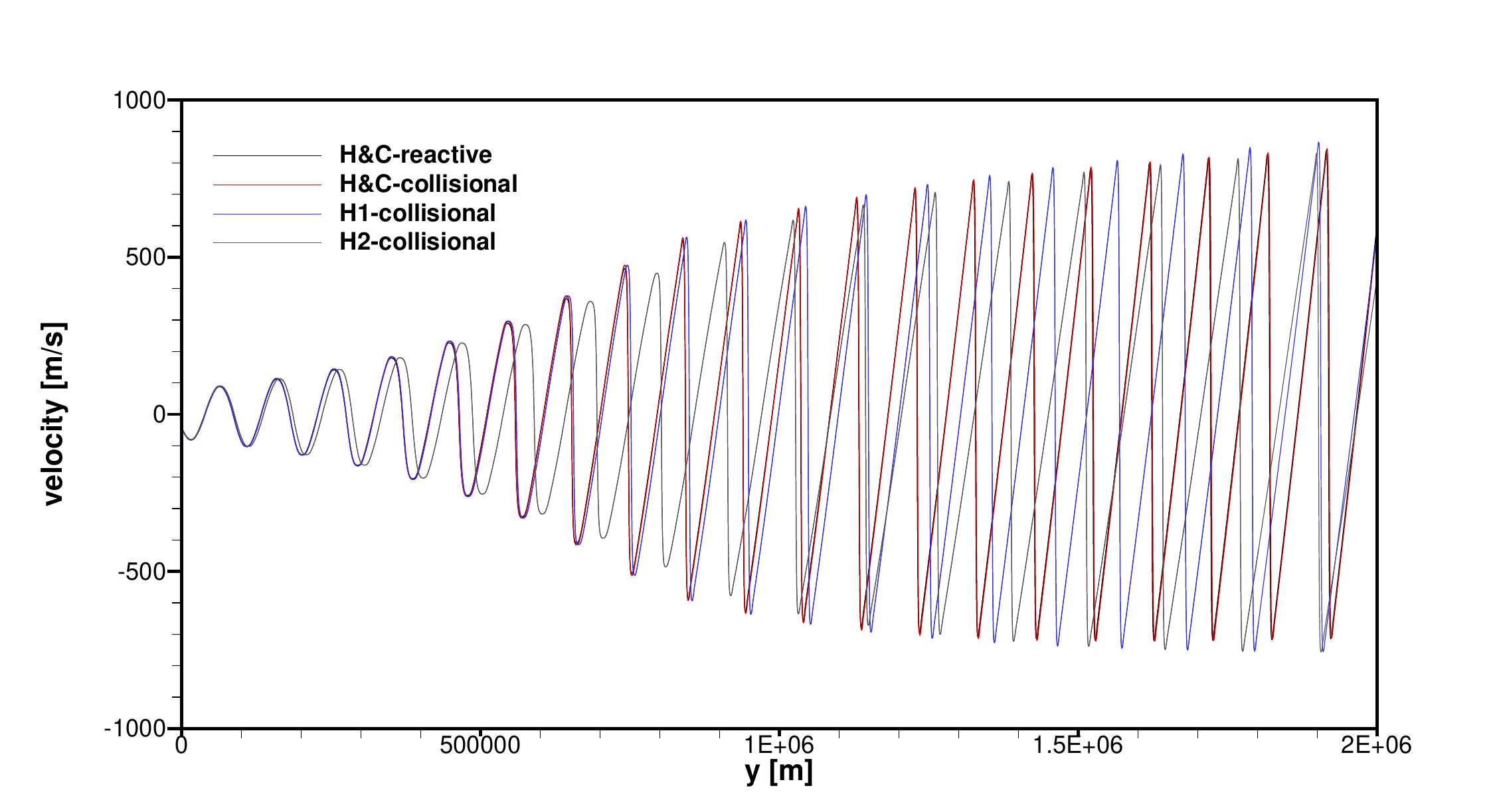}
\caption{Ion velocity profiles at $t=1200\;$s for wave period  $P=10\;$s. The velocity driver generates acoustic waves in media initialised by the density profiles shown in Fig.~\ref{fig:density}.  \label{fig:VelocityP10}}
\end{figure}

In Fig.~\ref{fig:VelocityP10}, the velocity profiles of the four numerical simulation results are shown. In general, the driver excites acoustic waves at the bottom boundary, and then the waves propagate upwardly, steepening to shocks higher up, since the density is exponentially decreasing in gravitationally stratified media.
It can be found that with higher ion density the waves propagate faster, because higher ionization fractions increase the sound speed in the present two-fluid plasmas. 
Moreover, we can also see that the amplitude increase slows down while reaching $y\approx 1\;$Mm, where the smooth acoustic waves start steepening to shocks, and further higher up, approximately constant amplitudes are found in all the velocity profiles.

\begin{figure*}
\gridline{\fig{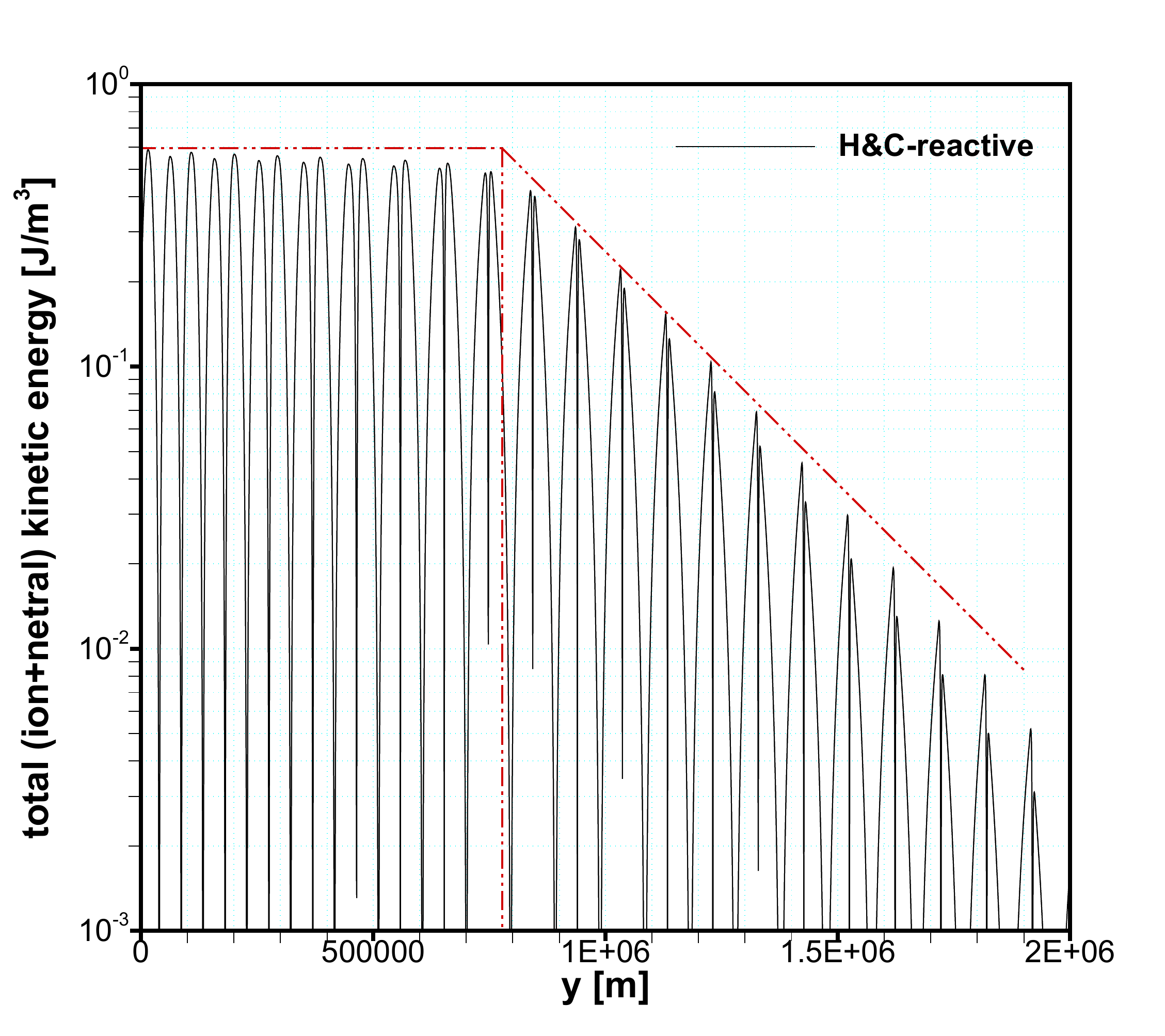}{0.5\textwidth}{(a)}
          \fig{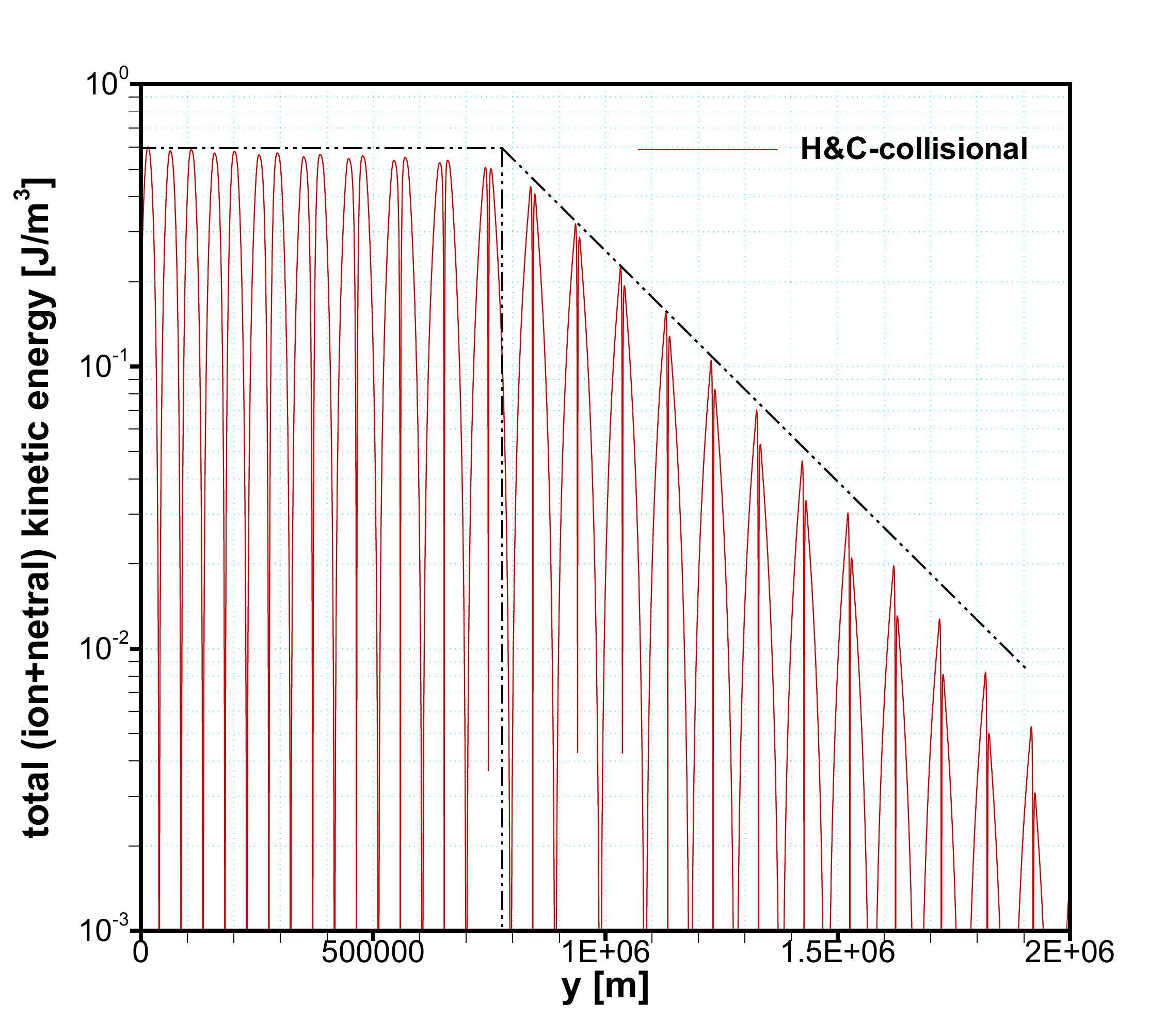}{0.5\textwidth}{(b)}
          }
\gridline{\fig{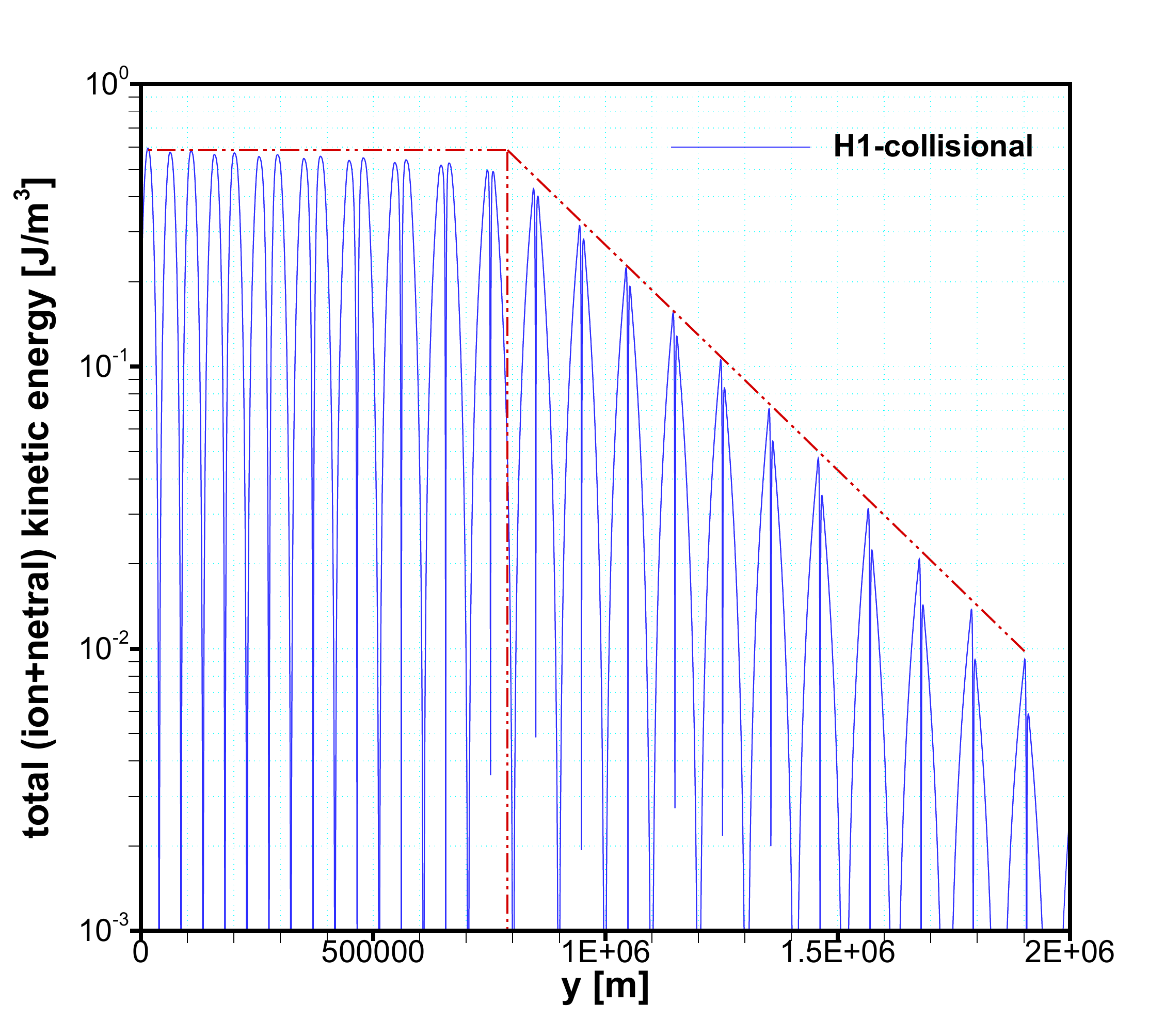}{0.5\textwidth}{(c)}
          \fig{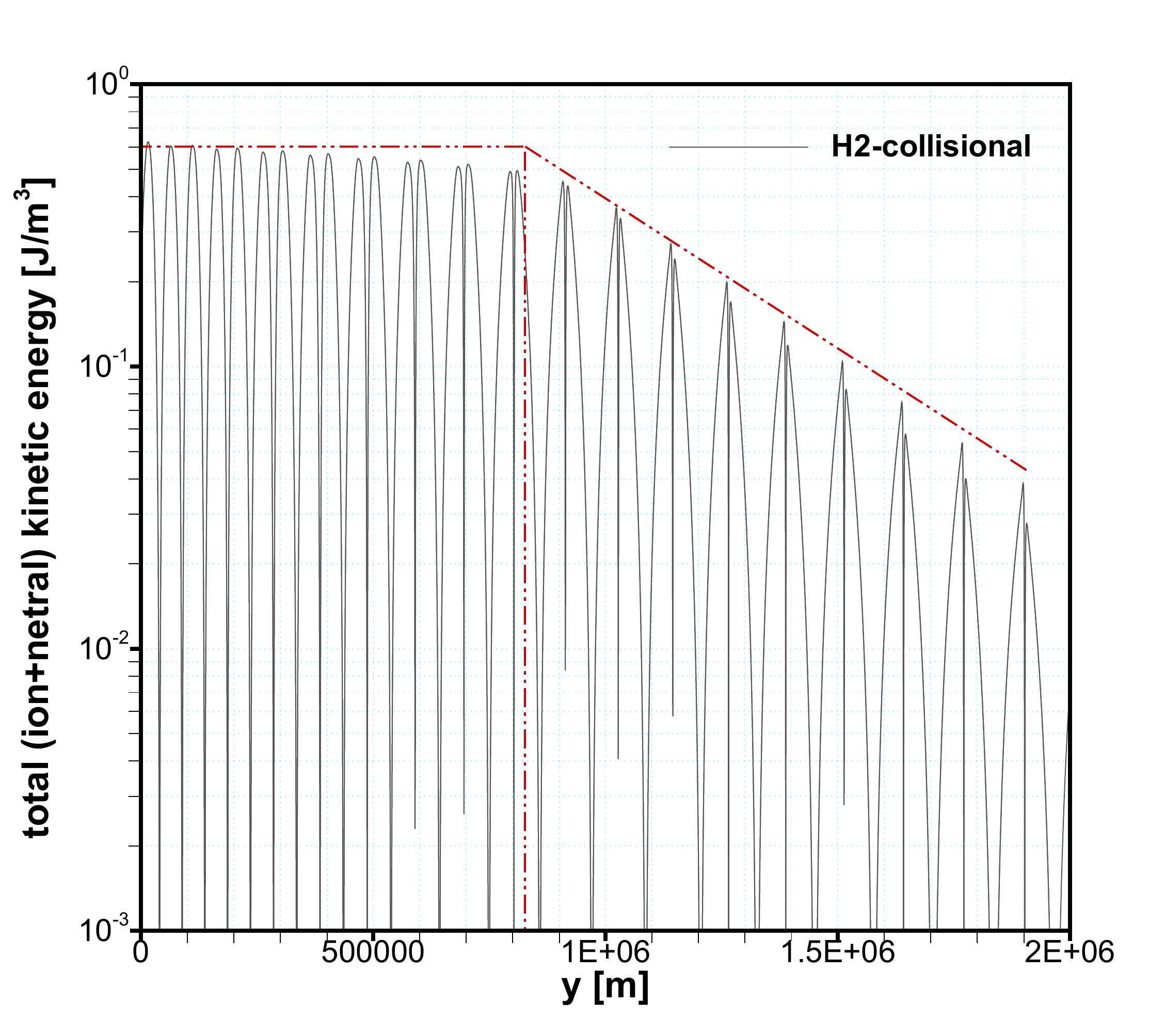}{0.5\textwidth}{(d)}
          }
\caption{Total (ion+neutral) kinetic energy profiles of vertical acoustic waves  at $t=1200\;$s for wave period  $P=10\;$s:
(a) reactive+collisional simulation with initial hydrostatic+chemical equilibrium, (b) collisional simulation with initial hydrostatic+chemical equilibrium, (c) collisional simulation with initial hydrostatic equilibrium, (d) collisional simulation with initial hydrostatic equilibrium \& high ion density. The vertical dashed-dotted lines indicate the approximate heights at which the strong nonlinear kinetic energy decays start, and the oblique dashed-dotted lines indicate the approximate decay rates of kinetic energy.
\label{fig:kineticE}}
\end{figure*}

However, the velocity profiles do not directly provide information about the energy deposition. Therefore, the kinetic energy of the waves (acoustic waves and shocks) is shown in Fig.~\ref{fig:kineticE}. It is interesting to see that the kinetic energy is almost constant in the lower regions, where shocks are not yet formed. Whereas, higher up, the kinetic energy decreases exponentially, indicating strong wave damping in the nonlinear regimes, where the acoustic waves develop to shocks. More specifically, with using the H\&C-profile, both the reactive and collisional simulations show significant wave damping, compared with two cases using (only) hydrostatic equilibrium initial density profiles. Moreover, as shown in Fig.~\ref{fig:kineticE}(d), with using the H2-profile, the wave damping is much slower, and the kinetic energy decay  starts at $y\approx 0.83\;$Mm, which is above the others. A minor issue should also be noted: while using the H\&C-profile, the energy decays are not strictly exponential, and the damping rate slightly increases in higher regions. In comparison, the H1- and H2-profiles not only have higher ion density, but also have slower ion density decreasing rates, resulting in exponential energy decays in almost the {entire nonlinear regime}. 

\begin{figure*}
\gridline{\fig{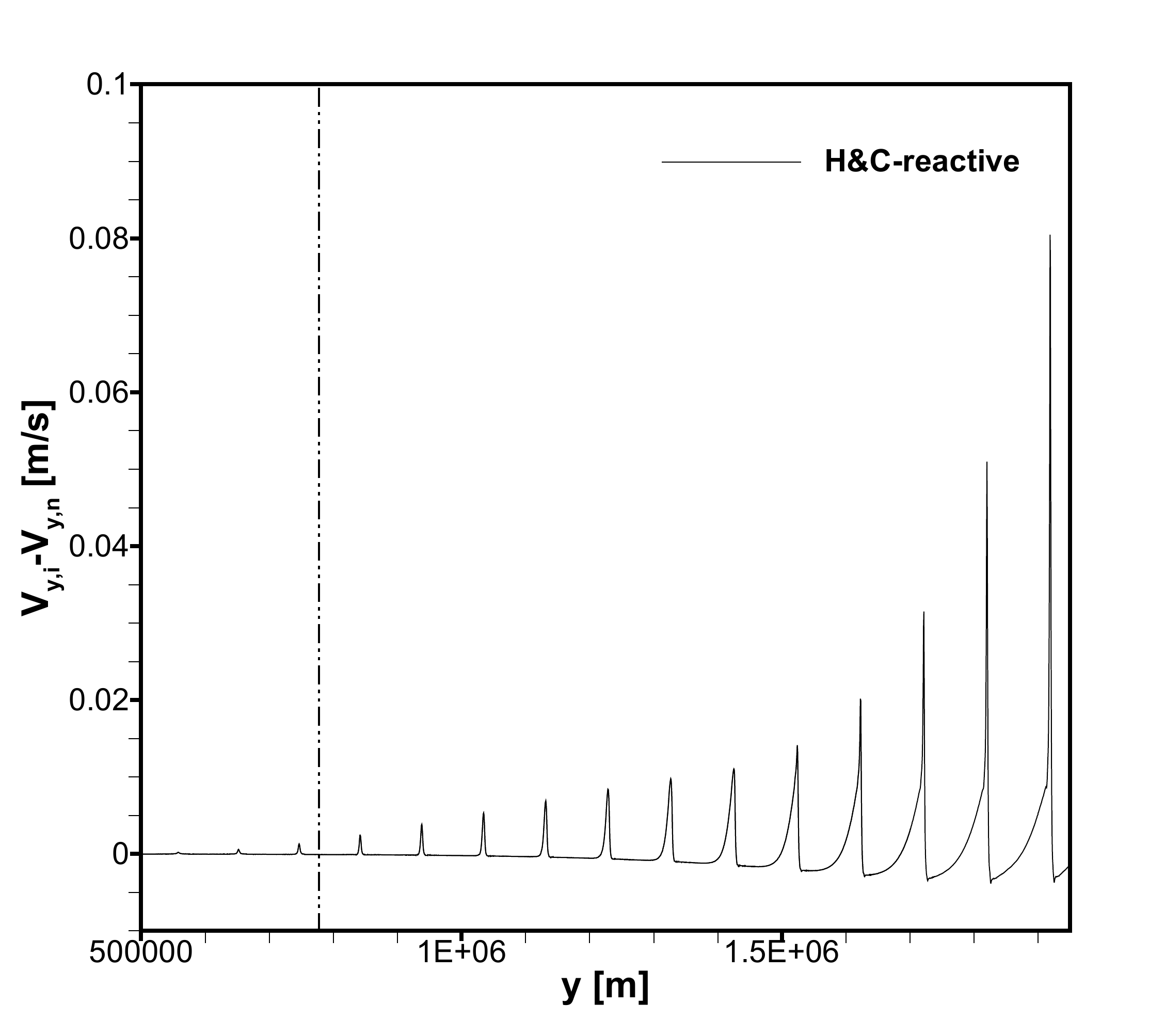}{0.5\textwidth}{(a)}
          \fig{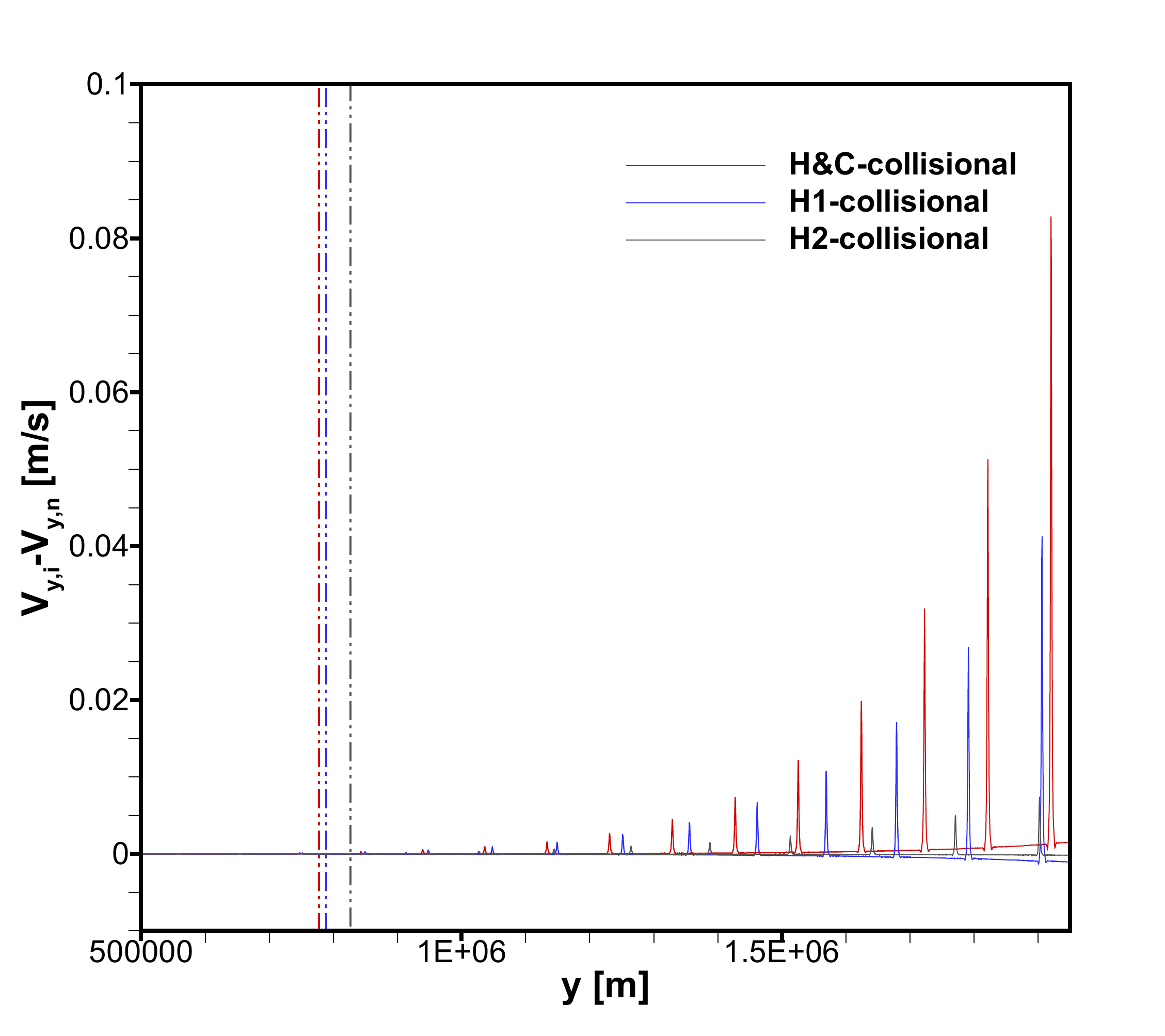}{0.5\textwidth}{(b)}
          }
\caption{Difference between vertical  components of ion and neutral velocities at $t=1200\;$s for wave period  $P=10\;$s:
(a) reactive+collisional simulation with initial hydrostatic+chemical equilibrium, (b) collisional simulations with initial hydrostatic+chemical equilibrium or hydrostatic equilibrium. The vertical dashed-dotted lines  indicate the approximate heights at which the strong nonlinear kinetic energy decays start, and each dashed-dotted line corresponds to the solid line of the same colour.
\label{fig:dV}}
\end{figure*}

We use the Euler equations for the fluid description, {in which viscosity is not taken into account,} and one may expect that the collisions are {an important} mechanism responsible for wave damping. The collision term is proportional to the square of the difference between the ion and neutral velocities. Fig.~\ref{fig:dV} illustrates $(v_{y,\text{i}}-v_{y,\text{n}})$, and the difference between the reactive simulation and the collisional simulations can be observed. All the collisional simulations show exponential increases in the ion and neutral velocity differences, and lower ion density leads to larger difference between the ion and neutral velocities. In particular, the peaks are at the shocks. {These behaviours seem to explain the kinetic energy damping shown in Fig.~\ref{fig:kineticE}: the strong energy decays start as soon as shocks occur and the fastest energy decay is found in the simulations having the lowest ion density profile. However, the velocity drift in the reactive simulation is larger than that of the collisional simulation using the same density profile. In particular, for  these two simulations using the H\&C-profile, the difference between them is more significant at shocks below $y \approx 1300\;$km and in regions between shocks, while the energy decays are almost the same.}

\begin{figure*}
\gridline{\fig{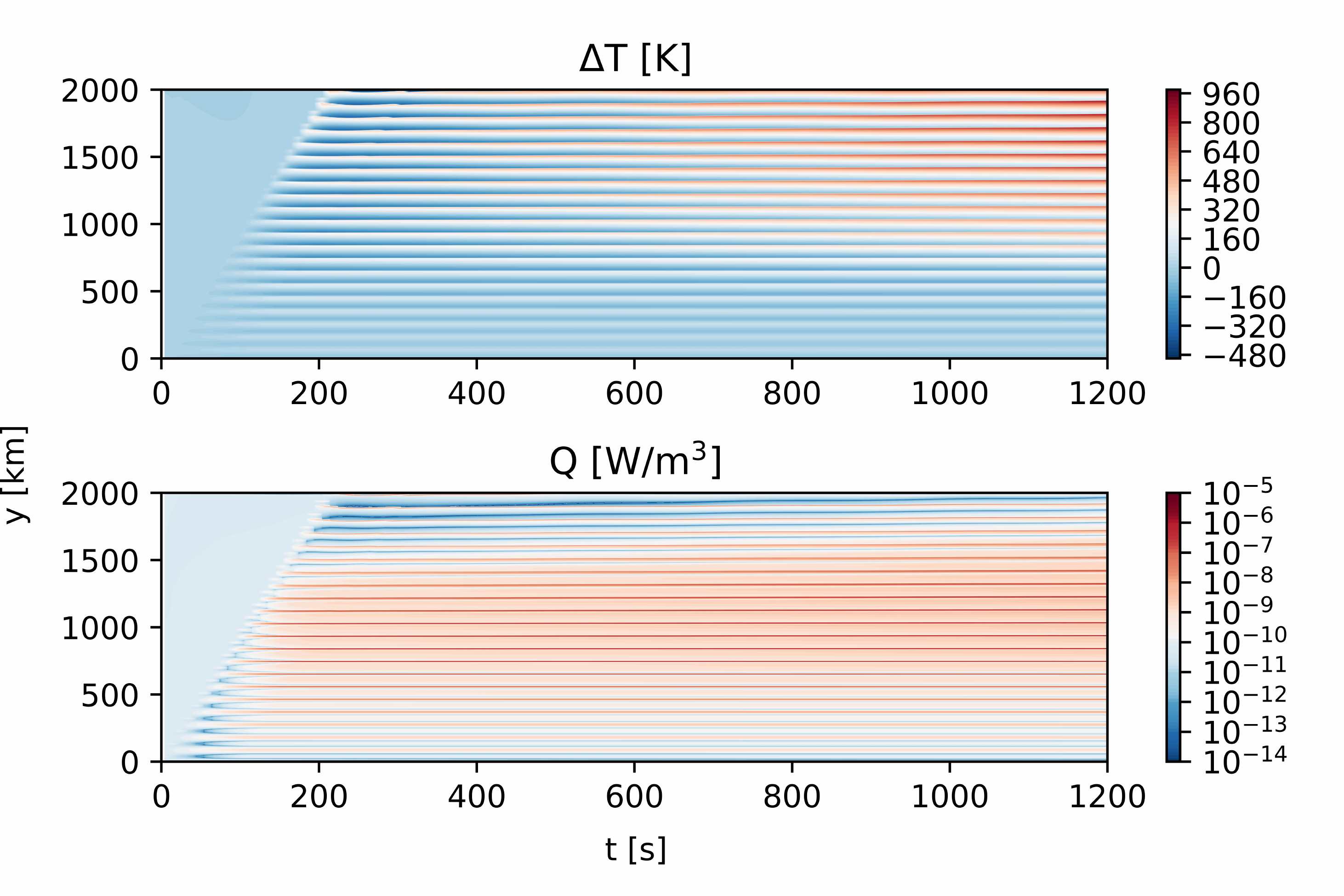}{0.5\textwidth}{(a)} 
          \fig{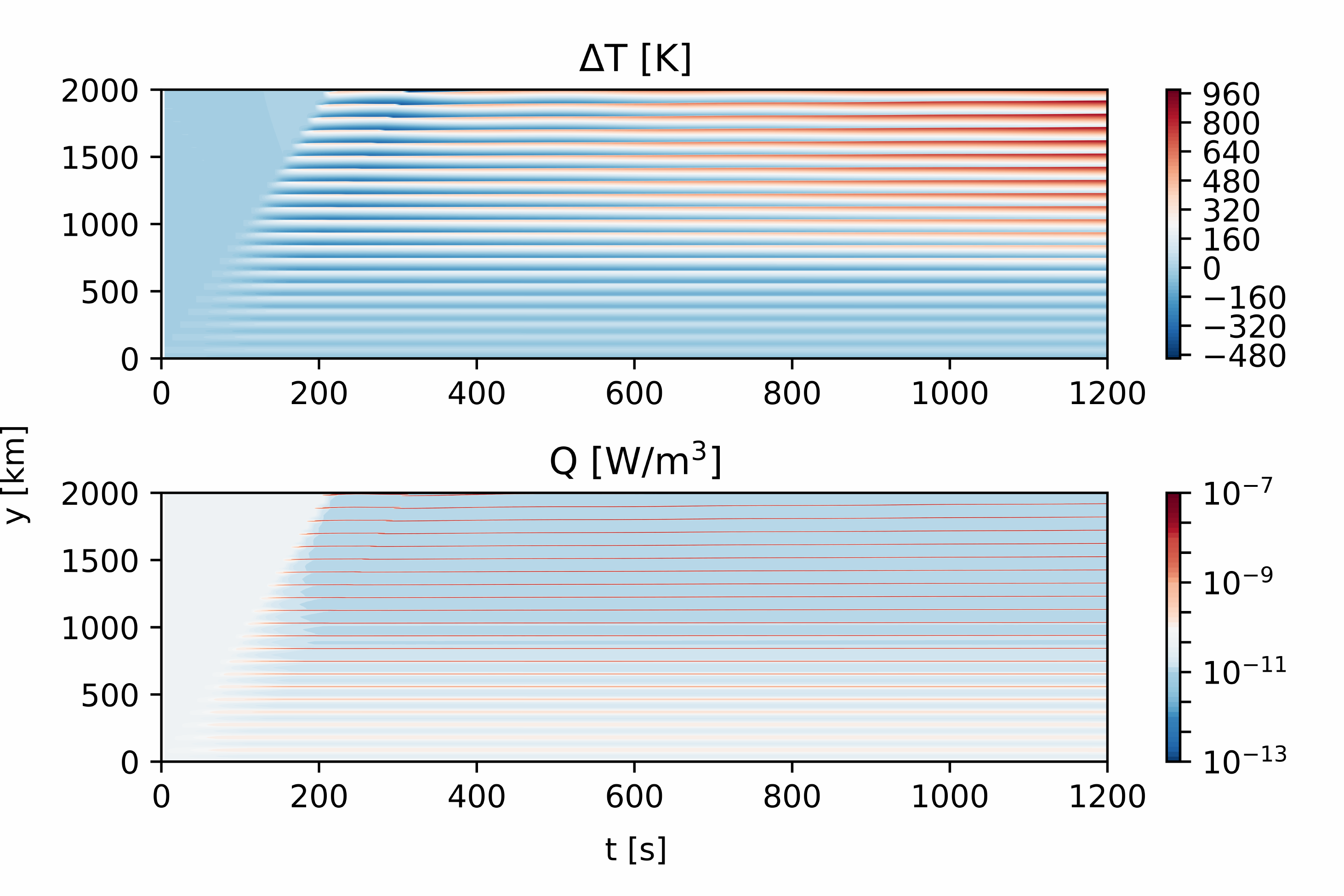}{0.5\textwidth}{(b)}
          } 
\gridline{\fig{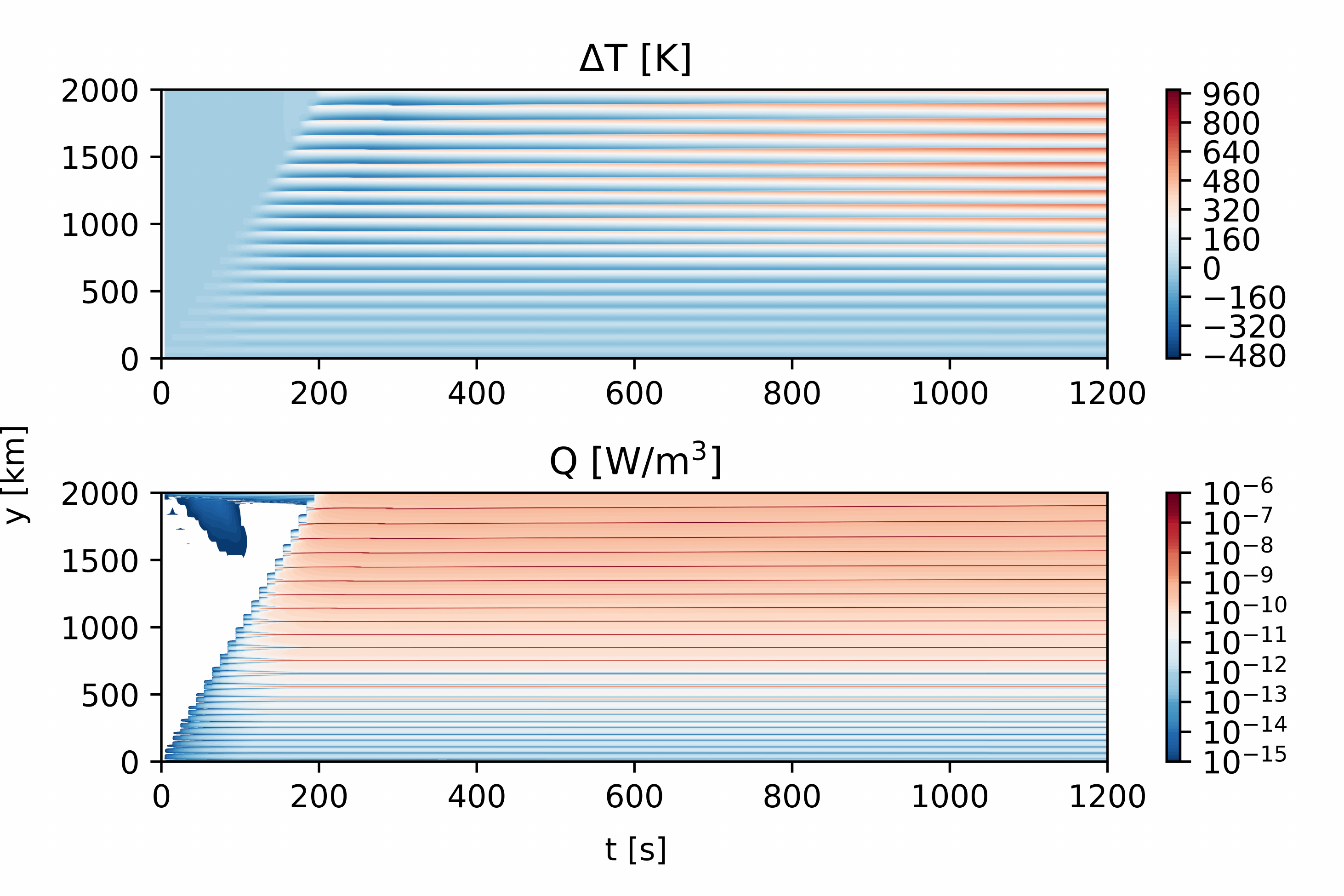}{0.5\textwidth}{(c)}
          \fig{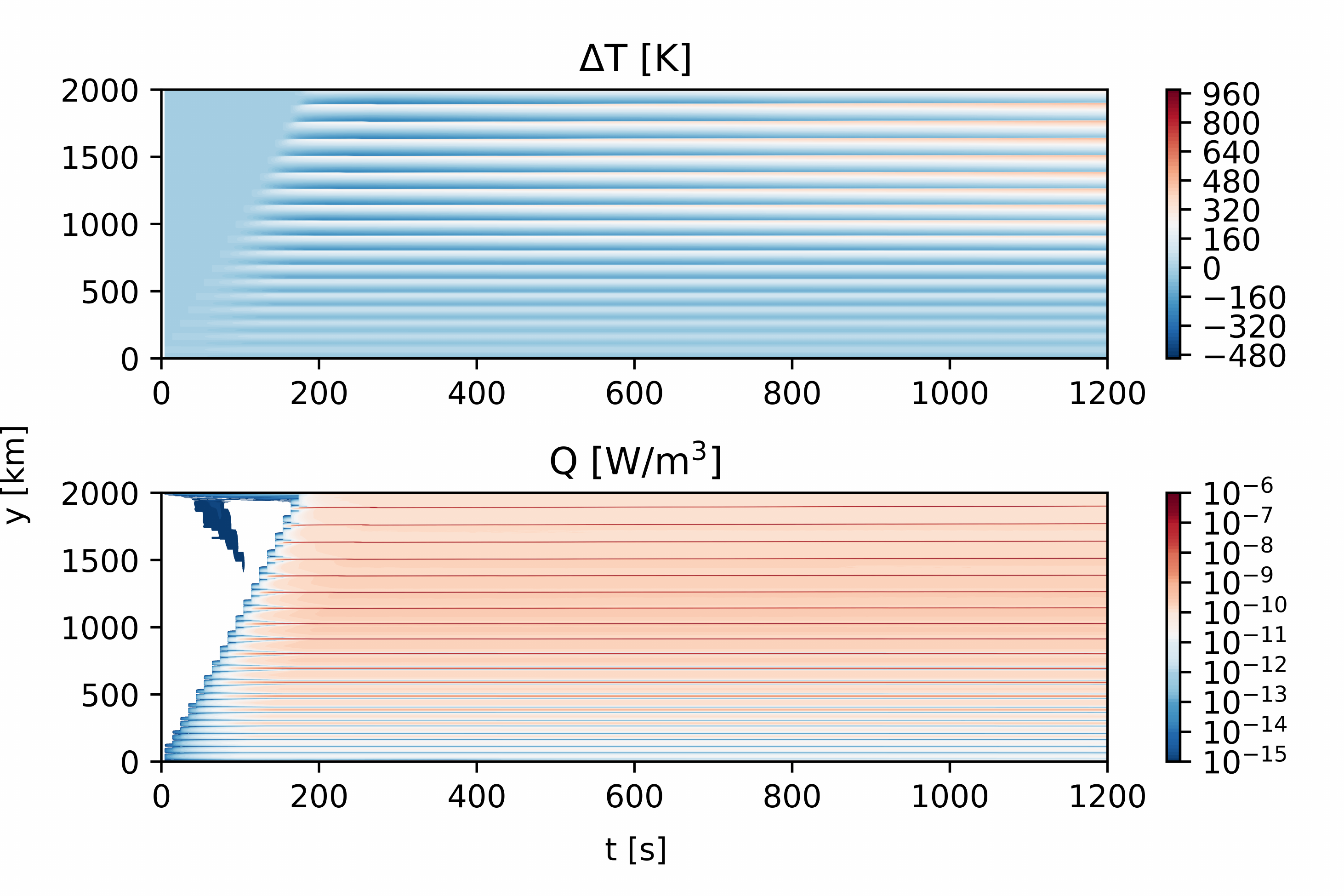}{0.5\textwidth}{(d)}
          } 
\caption{{Time-height plots of the net temperature increases and the collisional heating rates:
(a) reactive+collisional simulation with initial hydrostatic+chemical equilibrium, (b) collisional simulation with initial hydrostatic+chemical equilibrium, (c) collisional simulation with initial hydrostatic equilibrium, (d) collisional simulation with initial hydrostatic equilibrium \& high ion density. }
\label{fig:TAPlots_H}}
\end{figure*}

More detailed information of the {net} collisional (frictional) heating and the overall temperature increases are both shown in Fig.~\ref{fig:TAPlots_H}, {and the net collisional heating can be given as \citep{Laguna2017,PopescuBraileanu2019a}}
\begin{eqnarray}  
{Q}_{\text{collision}}=m_{\text{in}}n_{\text{i}}\nu_{\text{in}} (\mathbf{v}_{\text{n}}-\mathbf{v}_{\text{i}})^2,
\end{eqnarray}
\noindent {which includes the heating of ions and neutrals.} It can be found that the collisional heating of the collisional simulation using the H\&C-profile is significantly smaller (around two orders of magnitude) than that of the reactive simulation. Moreover, because the ion density is significantly higher in the H1- and H2-profile, the corresponding collisional heating is even stronger than that of the simulation using the H\&C-profile, which shows the most significant temperature increase. Therefore, the collisional heating should not be the major heating mechanism in the present simulations. Another and more essential evidence is the collisional frequency that can be calculated according to the plasma quantities. For the H\&C-profile, the neutral-ion collisional frequency reaches the minimum value $459\;$s$^{-1}$ at $y = 2000\;$km, which is still higher than the wave frequency. For the H1- or H2-profiles, since the ion density is significantly higher, the collisional frequencies are two or three orders of magnitude higher than that of the H\&C-profile. Although at shock wave fronts the  scale is greatly shortened, thus enhancing the decoupling of ions and neutrals and the collisional interactions, there is still not enough evidence supporting that the collisional interactions dominate the energy damping process. More specifically, even the maximum collisional heating at shock fronts is not sufficient to support the temperature increases being shown here. This is further discussed in the next subsection.

{In fact, the present heating effects can be explained by the classical shock heating theory. Although the present numerical results which include collisional and reactive effects in the two-fluid model cannot be fully reproduced by classical solutions such as the weak shock theory \citep{Ulmschneider1970,Ulmschneider1971,Ulmschneider1971_2,Stein1973,Jordan1973},  the method of characteristics \citep{Ulmschneider1977_2,Kalkofen1977,Ulmschneider1971} or the finite-volume solution for hydrodynamic  equations \citep{Kalkofen2010}, a qualitative explanation can be easily found. Firstly, because the energy dissipation only happens in the nonlinear regimes where shocks occur, the kinetic energy is constant below the heights of shock formation. Secondly, the height of shock formation is a function of sound speeds (and other parameters) \citep{Ulmschneider1971_2}, and in the present simulations with the given constant wave period and amplitude, higher sound speeds result in higher heights of shock formation. Finally, the shock dissipation process causes the kinetic energy decays. Of course, the collisional heating also contributes to the overall heating effects, but its contribution is small compared with the shock heating, which is shown in the next subsection in a more quantitative way.
}
 
\begin{figure}[ht!] 
 \centering
 \includegraphics[width=0.5\textwidth]{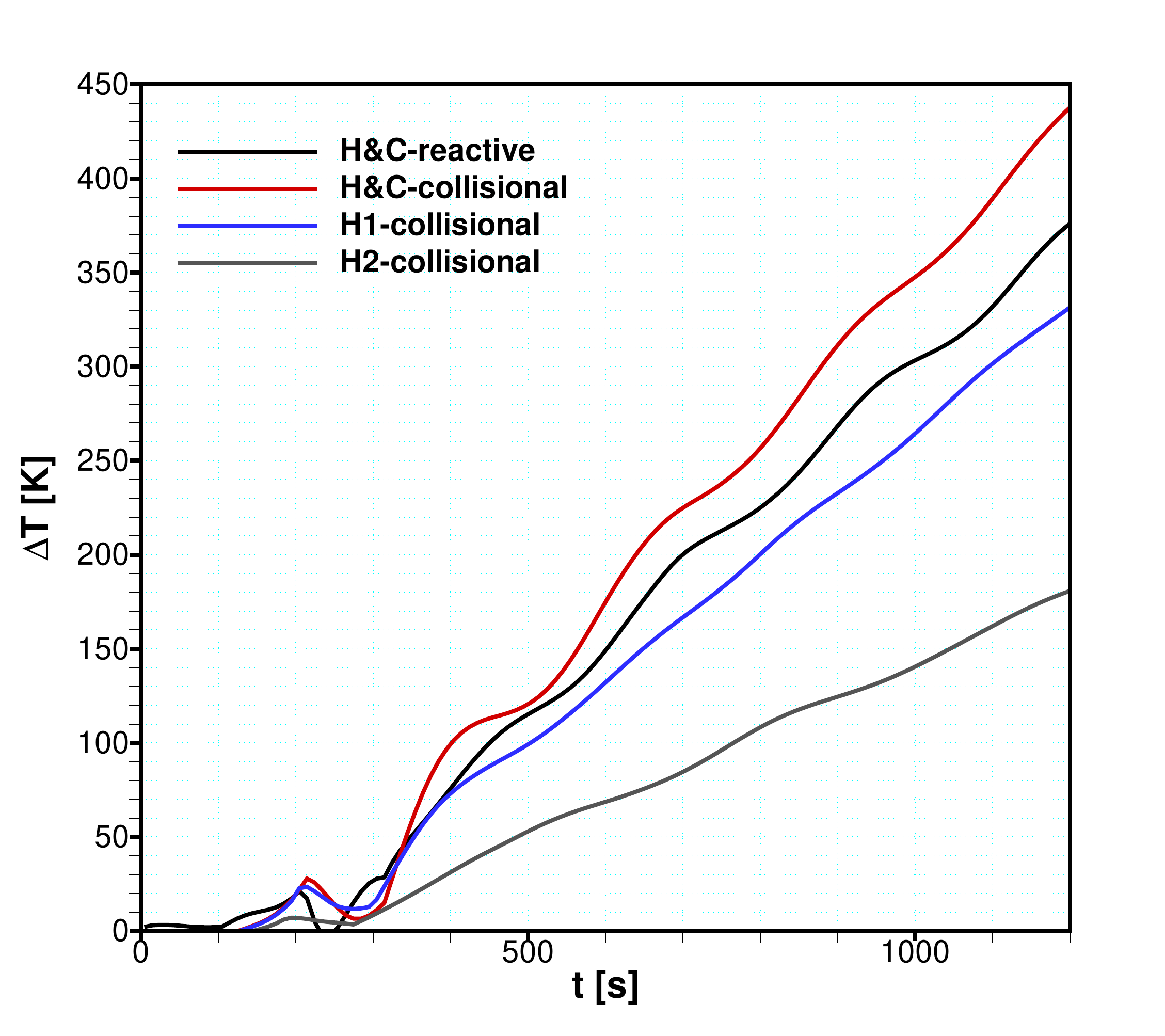}
\caption{{Spatially averaged ion temperature increases ($\Delta T$) over the region of ($1000$ km $ \le y \le 2000\;$km) versus time for wave period $P=10\;$s.}  \label{fig:TimeTemperatureP10}}
\end{figure}

\begin{figure*}
\gridline{\fig{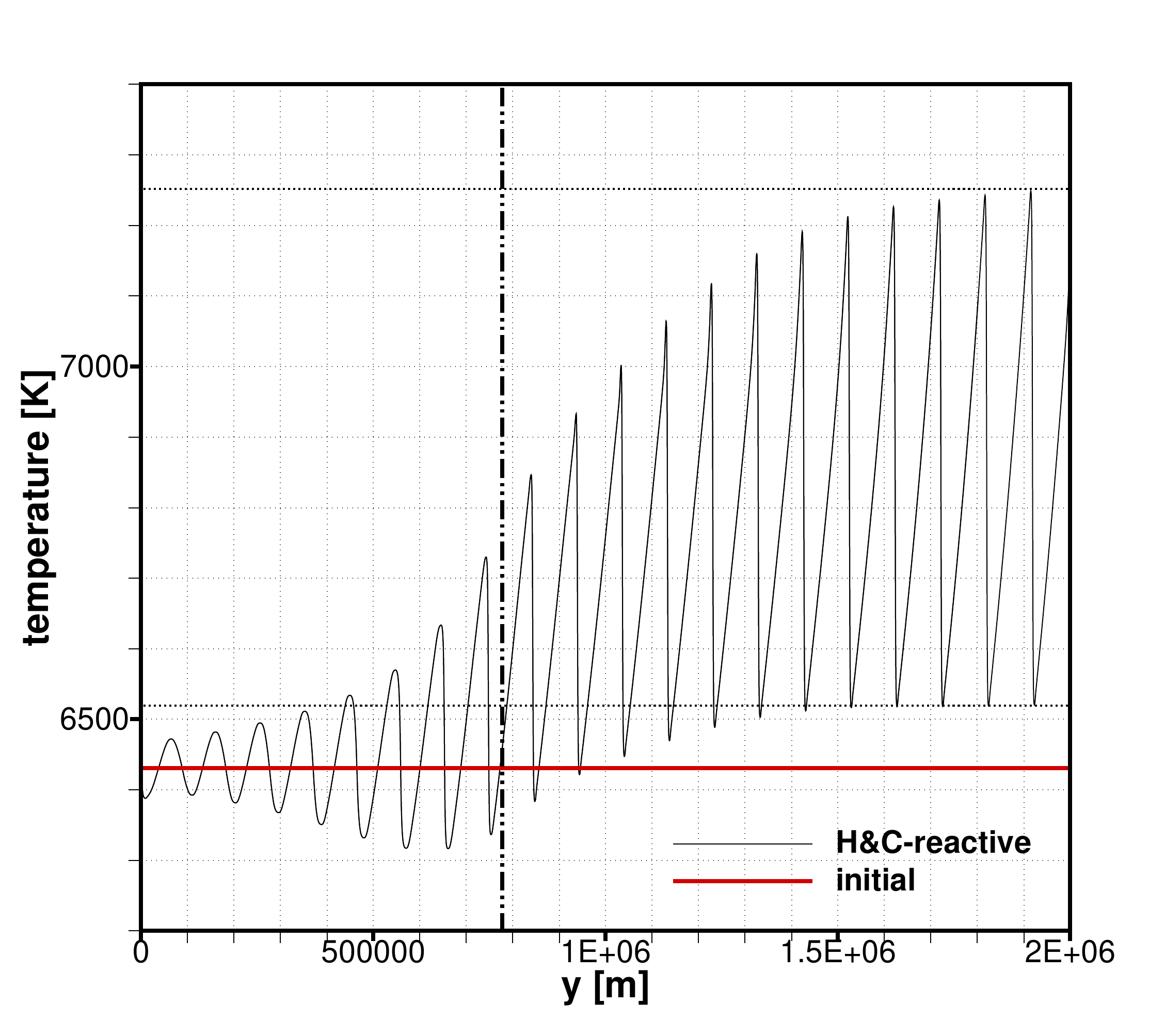}{0.5\textwidth}{(a)}
          \fig{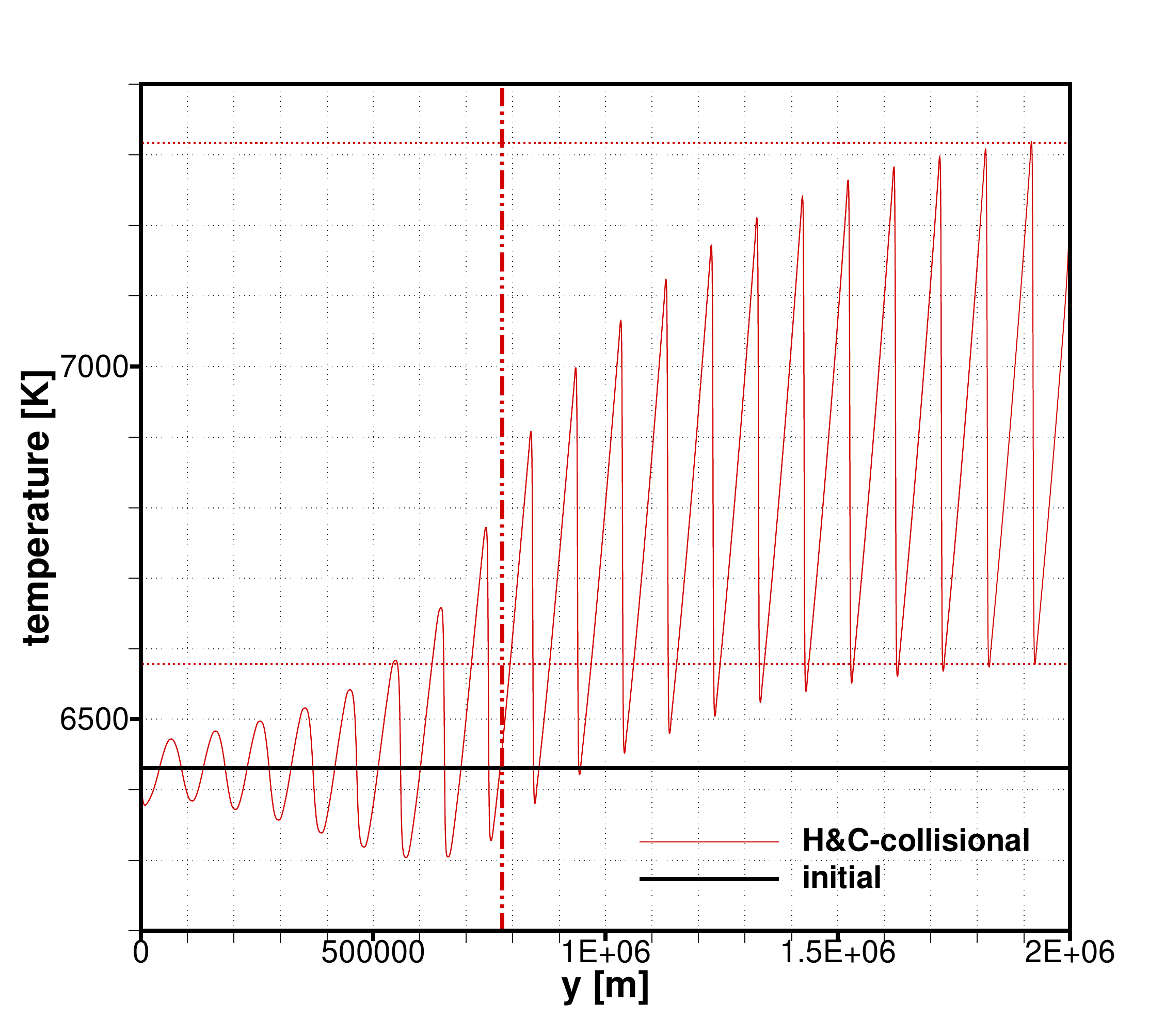}{0.5\textwidth}{(b)}
          }
\gridline{\fig{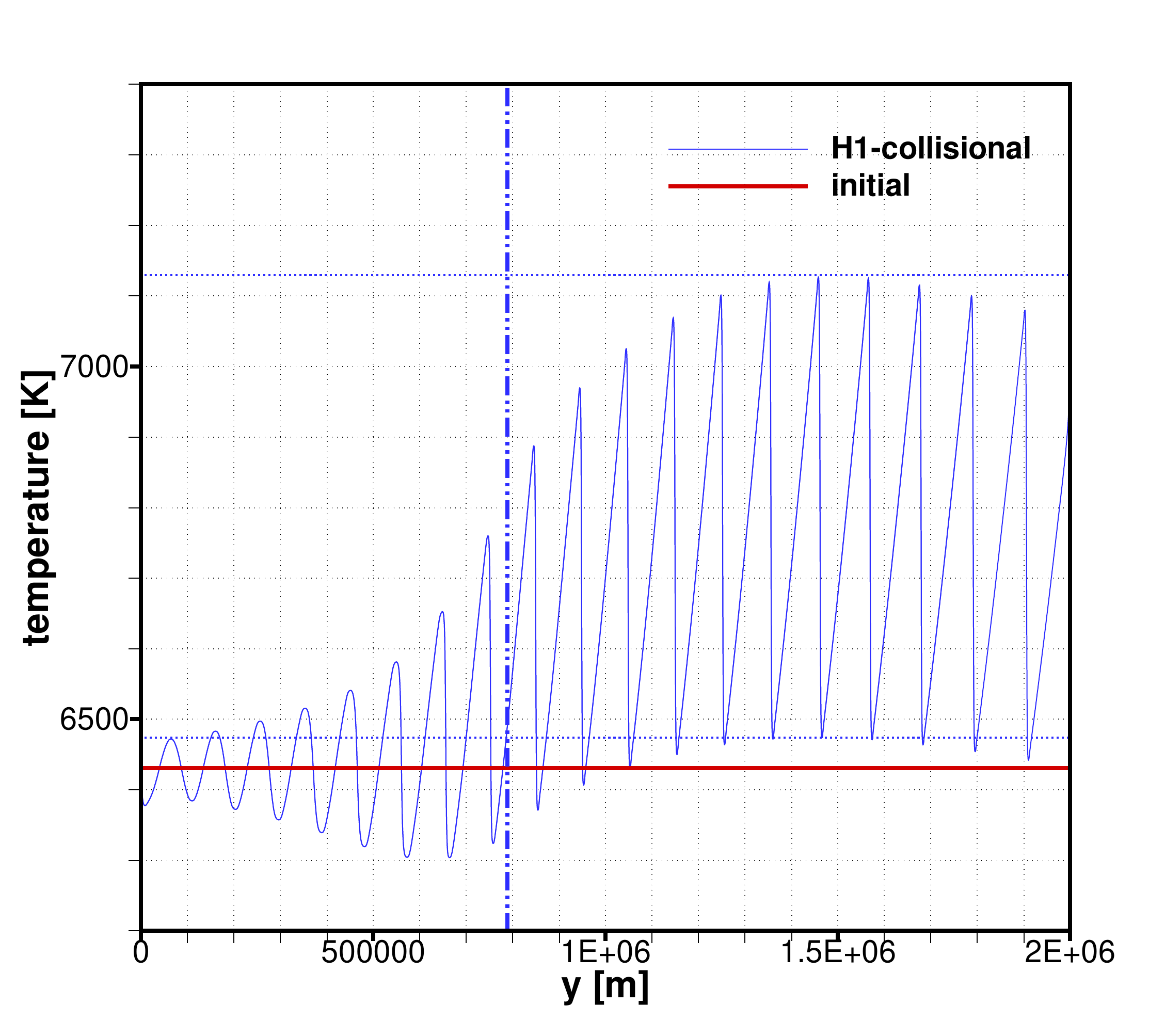}{0.5\textwidth}{(c)}
          \fig{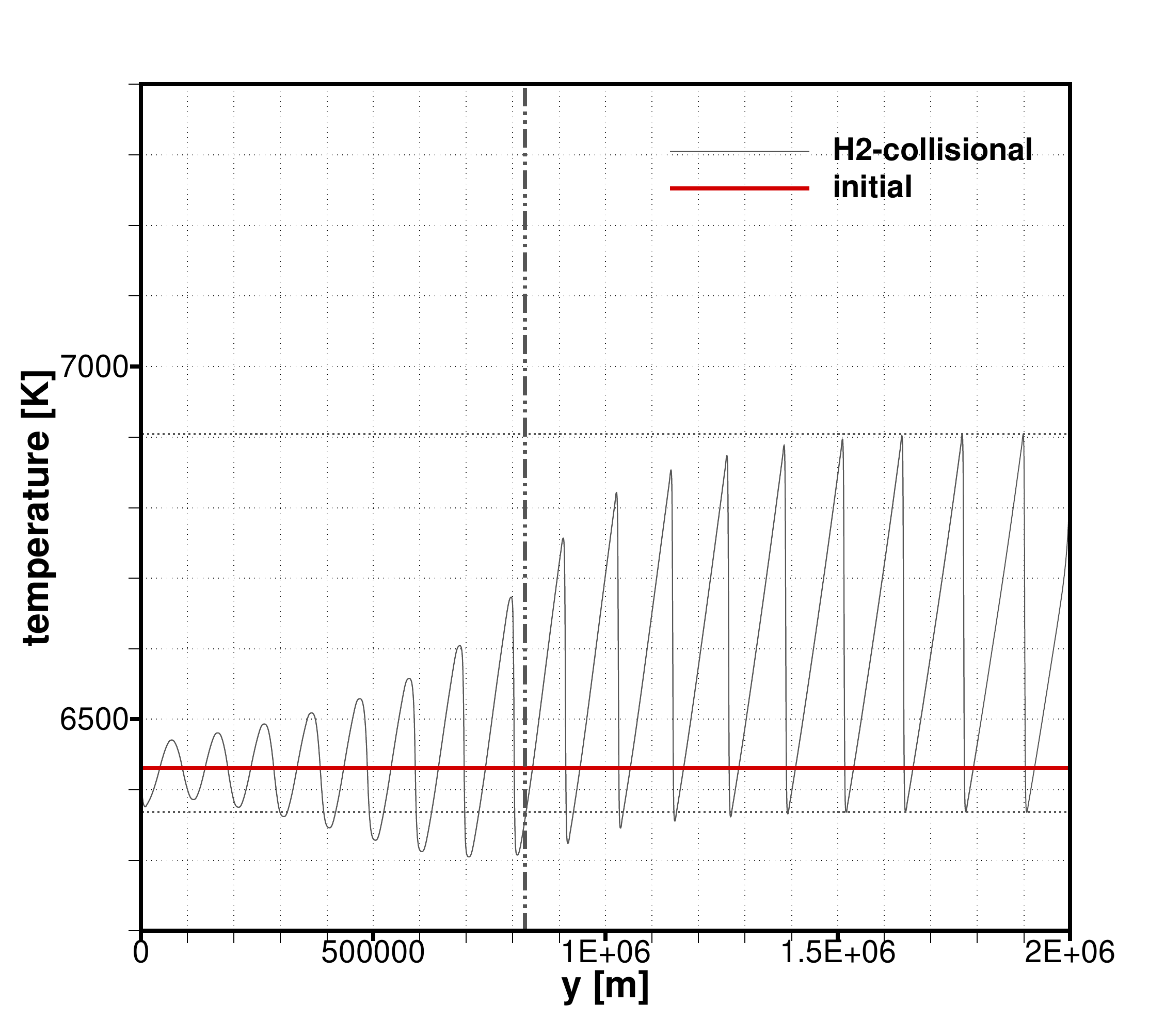}{0.5\textwidth}{(d)}
          }
\caption{Ion temperature profiles at $t=1200\;$s for wave period $P=10\;$s:
(a) reactive+collisional simulation with initial hydrostatic+chemical equilibrium, (b) collisional simulation with initial hydrostatic+chemical equilibrium, (c) collisional simulation with initial hydrostatic equilibrium, (d) collisional simulation with initial hydrostatic equilibrium \& high ion density. The vertical dashed-dotted lines indicate the approximate heights at which the strong nonlinear kinetic energy decays start.  The horizontal doted lines indicate the maximum/minimum temperatures of the hottest wave fronts, and thus the corresponding transient average temperature increase is $\Delta T=456\;$K (a), $518\;$K (b), $417\;$K (c), and $207\;$K (d).
\label{fig:TP10}}
\end{figure*}

Eventually, {the spatially averaged temperature increases over the region of ($1000$ km $ \le y \le 2000\;$km) versus time are shown in Fig.~\ref{fig:TimeTemperatureP10} and the temperature snapshots at $t=1200\;$s are shown} in Fig.~\ref{fig:TP10}. {It should be noted that the shock heating cannot be shown explicitly as the collisional heating which is described by a source term in the two-fluid equations, and thus only the temperature increments are shown here.} Obviously, the collisional simulation with the H\&C-profile shows the most significant heating, which leads to a maximum transient temperature increase of almost $900\;$K (more than twice  the increase when using the H2-profile). {The temperature variations also explain the net outflow shown in Fig.~\ref{fig:VelocityP10}, which should be caused by the pressure gradient resulting from the high temperature. In fact, with higher temperature, the net outflow velocity is also higher.} Moreover, it is interesting to see that the maximum heating occurs at different heights while using different density profiles. For instance, with using the H1-profile, the maximum temperature is found at around $y=1.5\;$Mm, but using the H\&C-profile leads to monotonously increasing temperature, as shown in Figs.~\ref{fig:TP10}(a) and (b). This should relate to the kinetic energy decays in Figs.~\ref{fig:kineticE}(a) and (b), which become faster at higher altitudes.

\begin{figure}[ht!]
 \centering
 \includegraphics[width=0.5\textwidth]{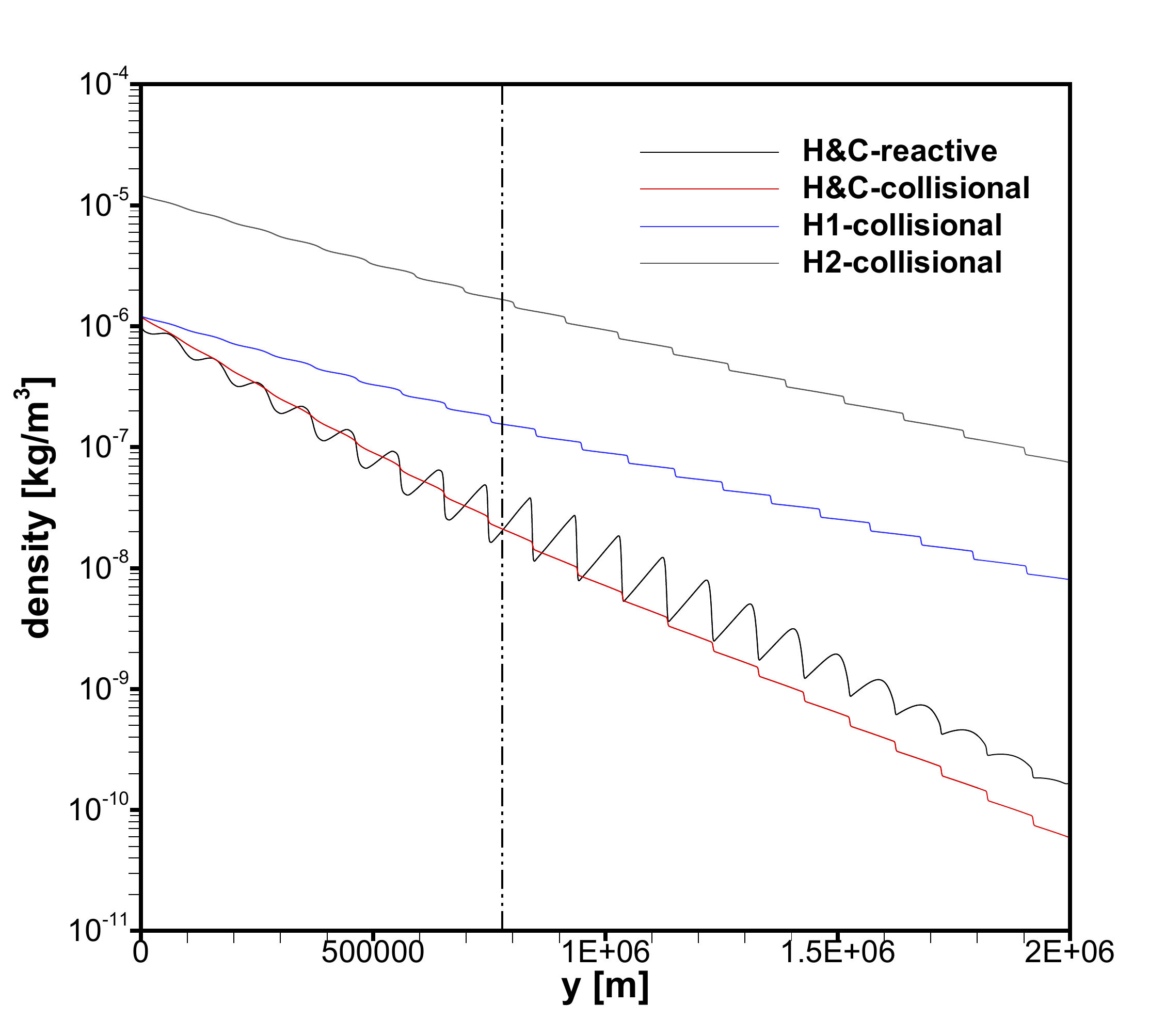}
\caption{Ion density profiles at $t=1200\;$s for wave period  $P=10\;$s.  The vertical dashed-dotted line  indicate the approximate height at which the strong nonlinear kinetic energy decay of the reactive+collisional simulation starts. \label{fig:densityP10}}
\end{figure}

While using the same density profile, the reactive simulation (Fig.~\ref{fig:TP10}(a)) shows  significant lower temperature compared with the collisional simulation (Fig.~\ref{fig:TP10}(b)), although they show similar kinetic energy decays. The explanation is actually straightforward: in the reactive simulation, while the temperature is increasing, the ionization process starts and requires a significant amount of energy, and thus it slows down the heating process {\citep{Stein1972,Stein1973}}. In Fig.~\ref{fig:densityP10}, the ion density profiles further support this explanation. We can find that the resulting ion density of the reactive simulation is higher than that of the collisional simulation, due to the ionization process. {More importantly, taking into account the ionization and recombination processes also enhances the decoupling between ions and neutrals and the collisional heating, as shown in Fig.~\ref{fig:dV} and Fig.~\ref{fig:TAPlots_H}, respectively. This enhancement of decoupling occurs probably (partly) due to the momentum exchanges caused by the ionization and recombination processes. In the meantime, the ionization process consumes neutrals and produces more ions, locally breaking the hydrostatic equilibrium, and changing the other properties of the plasma, e.g. the sound speed. Therefore, the resulting local imbalance may also enhance the decoupling.}

\subsection{Different Wave periods of Driven Acoustic Waves}

In the previous subsection, we have {numerically investigated the acoustic wave propagation while considering different density profiles and a given wave period ($10\;$s), showing significantly different wave damping and heating efficiencies.  It is also known that the damping and heating efficiencies are strongly affected by 
the wave periods. For instance, \citet{Ulmschneider1977,Kalkofen1977,Ulmschneider1977_2} showed that imposing a longer acoustic wave period leads to a higher height of shock formation, and more importantly, long-period shock waves will also have increasing dissipation with height as the density is exponentially decreasing. Recently, \citet{Kuzma2019} and \citet{PopescuBraileanu2019} have further investigated the heating effects of acoustic waves and magneto-acoustic waves with different wave periods, by using two-fluid numerical modeling. It is also} interesting to revisit the effects of different wave periods while taking into account the initial chemical equilibrium and the reactive interactions between ions and neutrals.
In this subsection, we use the H\&C-profile for initializing all the density distributions, and then impose velocity drivers having different wave periods, {as introduced in subsection \ref{sec:BC_IC}}. Both collisional and reactive(+collisional) simulations are performed.

\begin{figure}[ht!]
 \centering
 \includegraphics[width=0.7\textwidth]{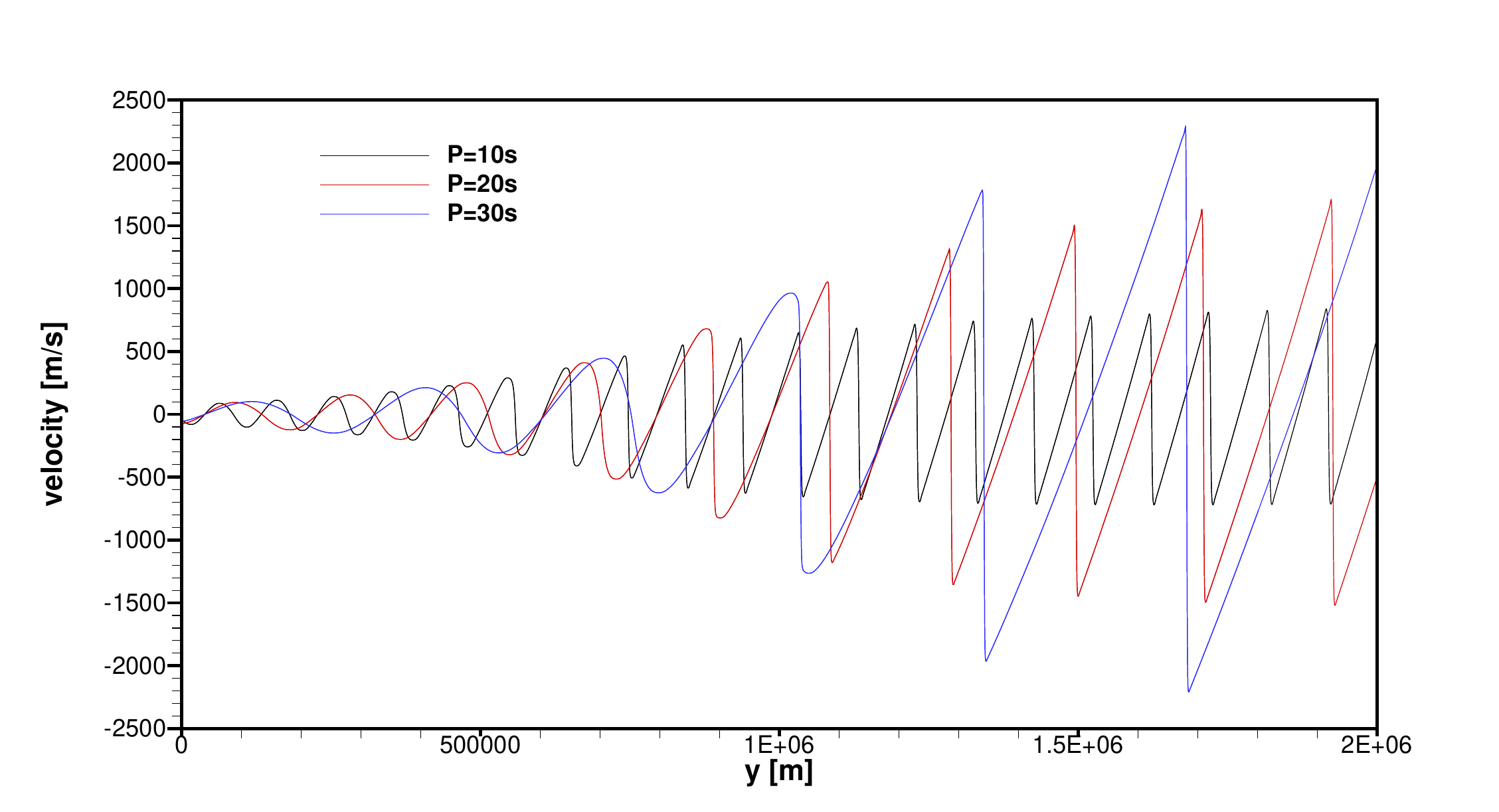}
\caption{Ion velocity profiles of reactive+collisional simulations at $t=1200\;$s. \label{fig:VelocityReactive}}
\end{figure}

\begin{figure}[ht!]
 \centering
 \includegraphics[width=0.7\textwidth]{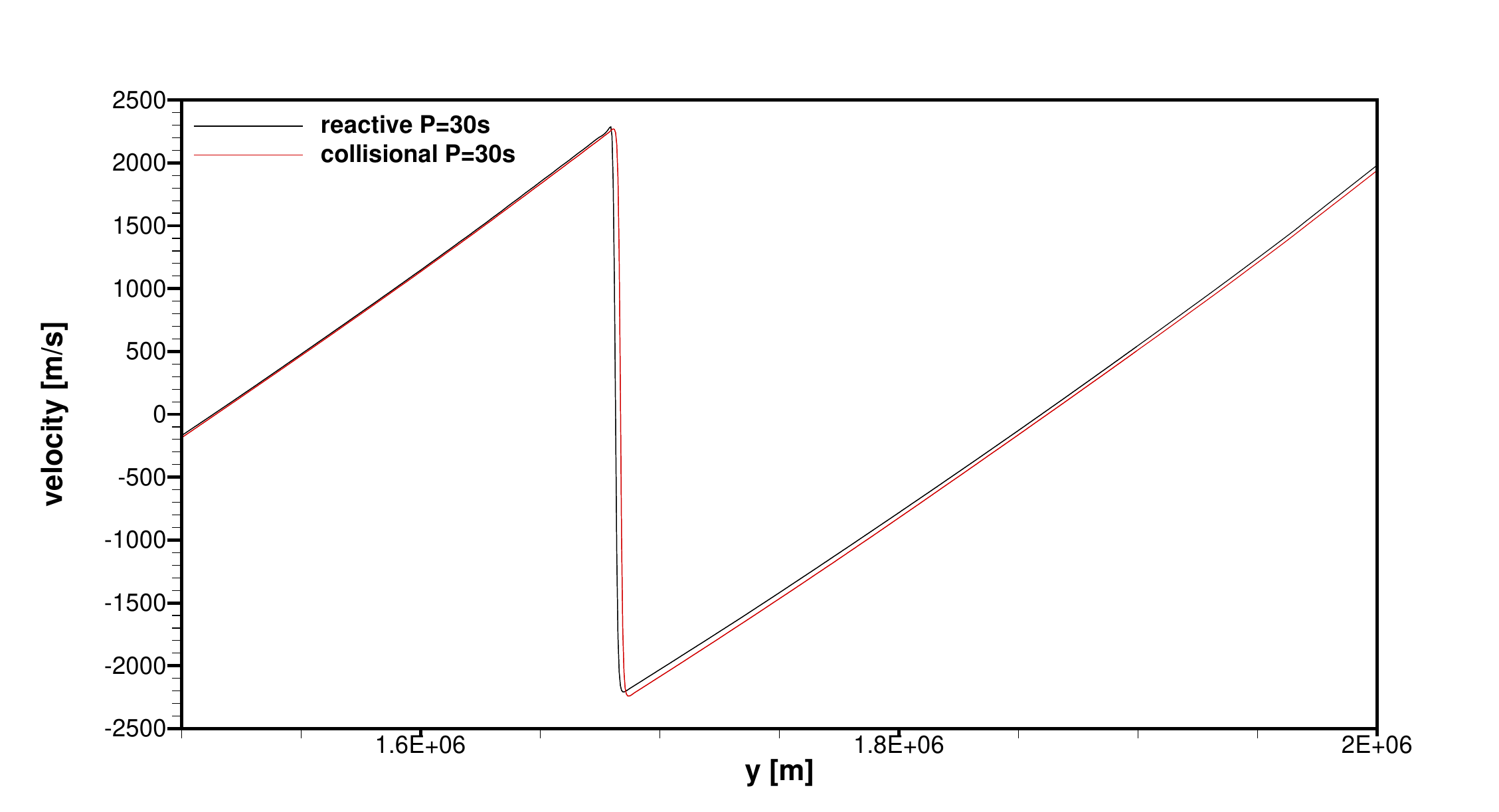}
\caption{Ion velocity profiles at $t=1200\;$s for wave period $P=30\;$s, showing the difference between the reactive+collisional simulation and the  collisional simulation.  \label{fig:P30VComparison}}
\end{figure}

\begin{figure*}
\gridline{\fig{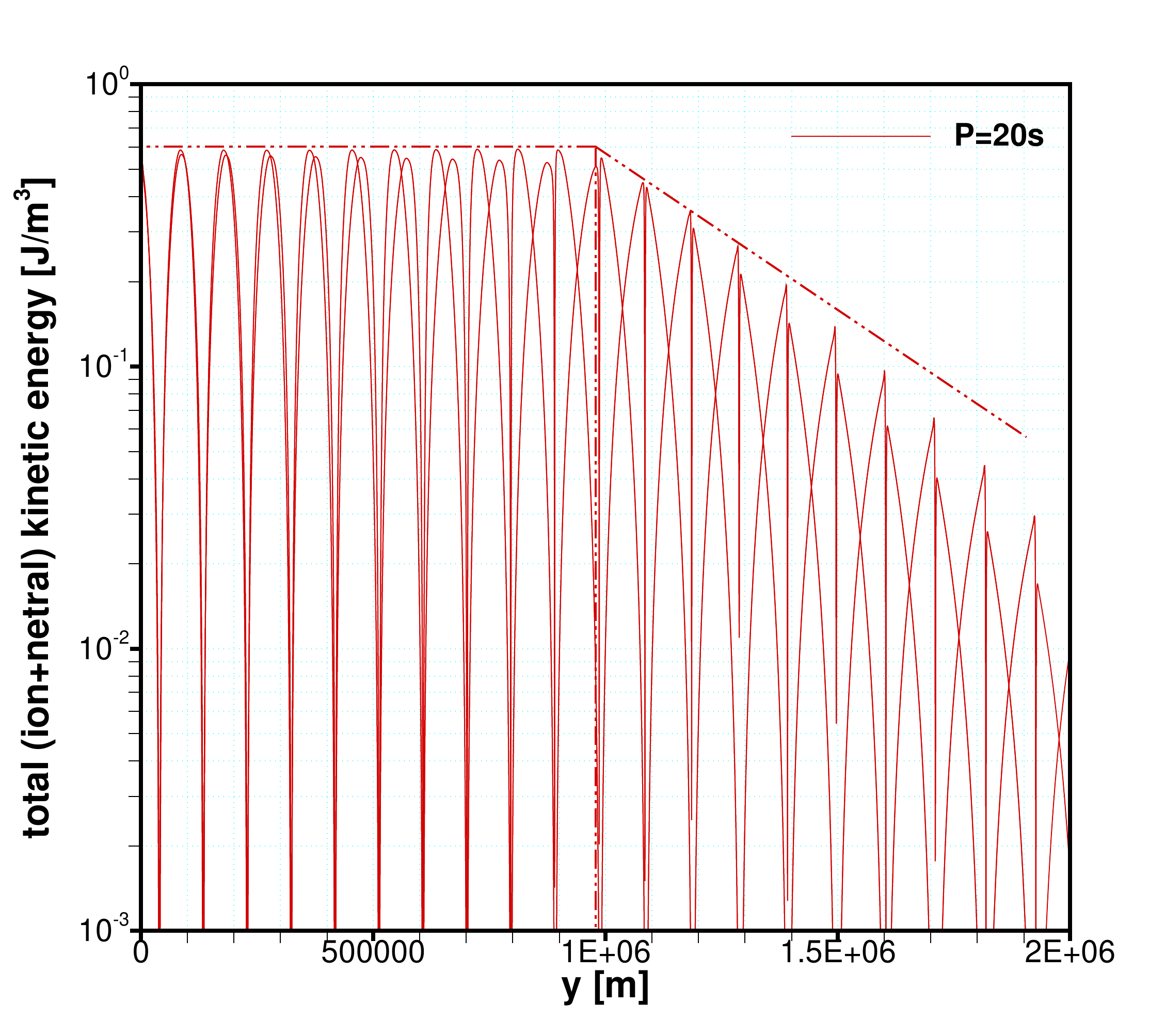}{0.5\textwidth}{(a)}
          \fig{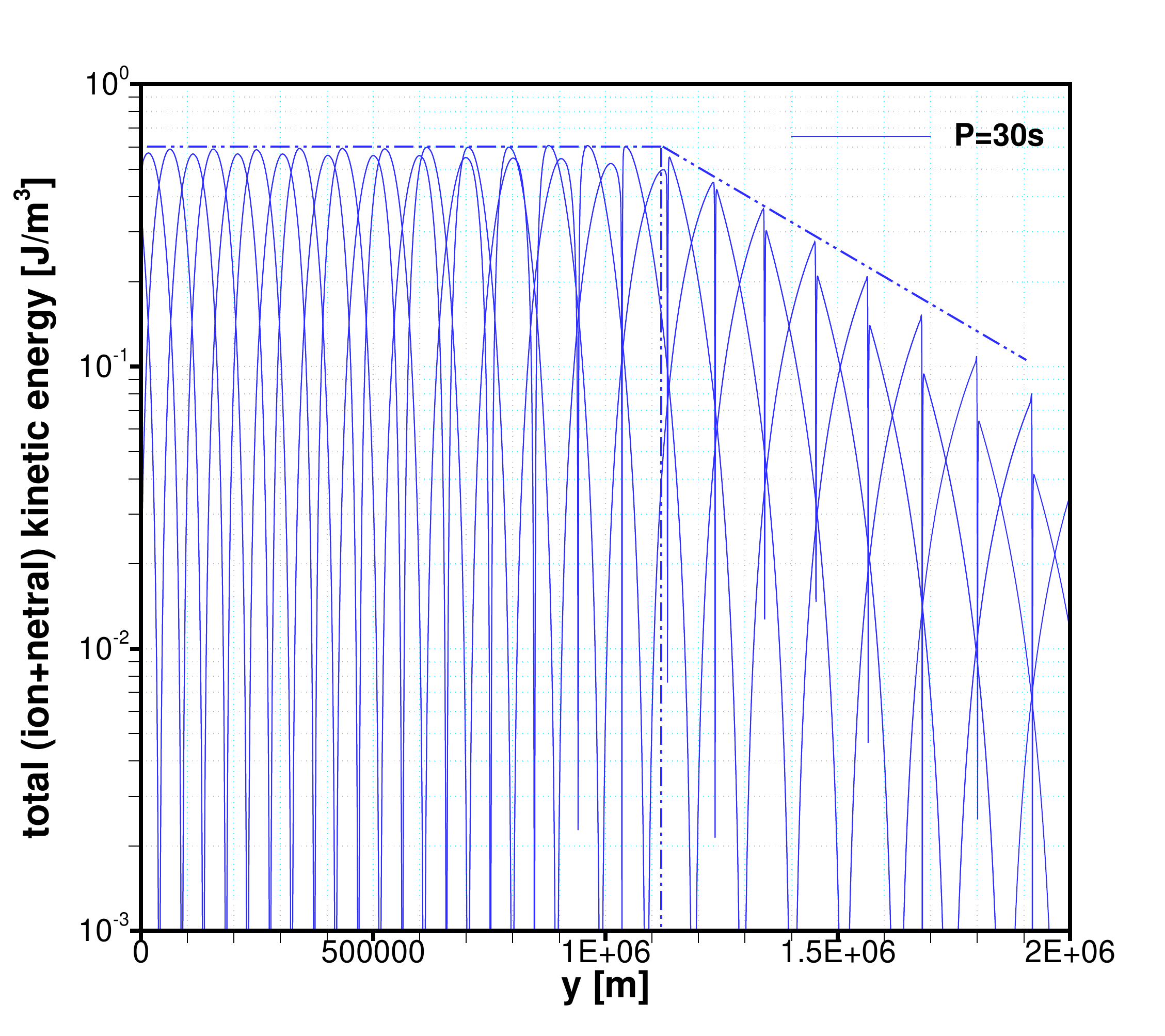}{0.5\textwidth}{(b)}
          }
\caption{Total (ion+neutral) kinetic energy profiles of vertical acoustic waves of reactive+collisional simulations.  Two snapshots   at $t=1200\;$s and $t=1210\;$s (a), and three snapshots   at $t=1190\;$s, $t=1200\;$s and $t=1210\;$s (b), are shown together to better provide the energy decay profiles.
The vertical dashed-dotted lines indicate the approximate heights at which the strong nonlinear kinetic energy decays start, and the oblique dashed-dotted lines indicate the approximate decay rates of kinetic energy.
\label{fig:KineticP2030Reactive}}
\end{figure*}

Firstly, Fig.~\ref{fig:VelocityReactive} shows the ion velocity profiles of the reactive simulations. However, the velocity profiles of the collisional simulations are not shown here, since they are very close to the reactive simulation results. In Fig.~\ref{fig:P30VComparison}, a comparison is provided, showing that with the same wave period $P=30\;$s, the difference between the amplitudes of the velocity profiles of the reactive and collisional simulations is rather small.
In Fig.~\ref{fig:VelocityReactive}, one may directly observe that a longer wave period leads to a larger wave amplitude. The maximum wave amplitudes of the present simulations seem to be approximately proportional to the wave periods. Moreover, again, more information can be found in the kinetic energy profiles, as shown in Fig.~\ref{fig:KineticP2030Reactive}, and only the results of the reactive simulations are shown here since the kinetic energy profiles of the collisional simulations are again similar as long as the same density profile and the same wave period are adopted. By observing Figs.~\ref{fig:KineticP2030Reactive} and \ref{fig:kineticE}(a), we can further confirm that indeed the strong damping starts at lower heights for shorter wave periods, and detailed information is shown in Table~\ref{tab:Slopes}, where several collisional simulation results are not shown 
{because of the similarity. In general, the basic behaviours found here are similar to the description of 1D acoustic/shock wave propagation provided by \citet{Ulmschneider1977_2} and others.}


\begin{deluxetable*}{ccccccc}
\tablenum{1}
\tablecaption{The heights at which strong nonlinear damping starts and the slopes of the kinetic energy decays\label{tab:Slopes}}
\tablewidth{0pt}
\tablehead{
\colhead{Source terms} & \colhead{Density profile} & \colhead{Wave period [s]} & \colhead{Height [Mm]} &
\colhead{Slope}
}
\startdata
Collision           & H\&C  & 10 & 0.78 & -4.7  \\
Collision           & H1    & 10 & 0.79 & -4.6    \\
Collision           & H2    & 10 & 0.83 & -3.2    \\
Reaction+Collision  & H\&C  & 10 & 0.78 & -4.7  \\
Reaction+Collision  & H\&C  & 20 & 0.98 & -3.5   \\
Reaction+Collision  & H\&C  & 30 & 1.1  & -3.3 \\
\enddata
\tablecomments{The heights are approximately provided, since there are not exact boundaries separating different regions. {The slopes are also approximately calculated for the kinetic energy decays (the oblique dashed-dotted lines in Figs.~\ref{fig:kineticE} and \ref{fig:KineticP2030Reactive}) in log-log scale, while the strong nonlinear damping rates are not exactly constant at different altitudes, particularly if the H\&C-profile is applied (Figs.~\ref{fig:kineticE}(a), (b) and \ref{fig:KineticP2030Reactive}).}}
\end{deluxetable*}

\begin{figure*}
\gridline{\fig{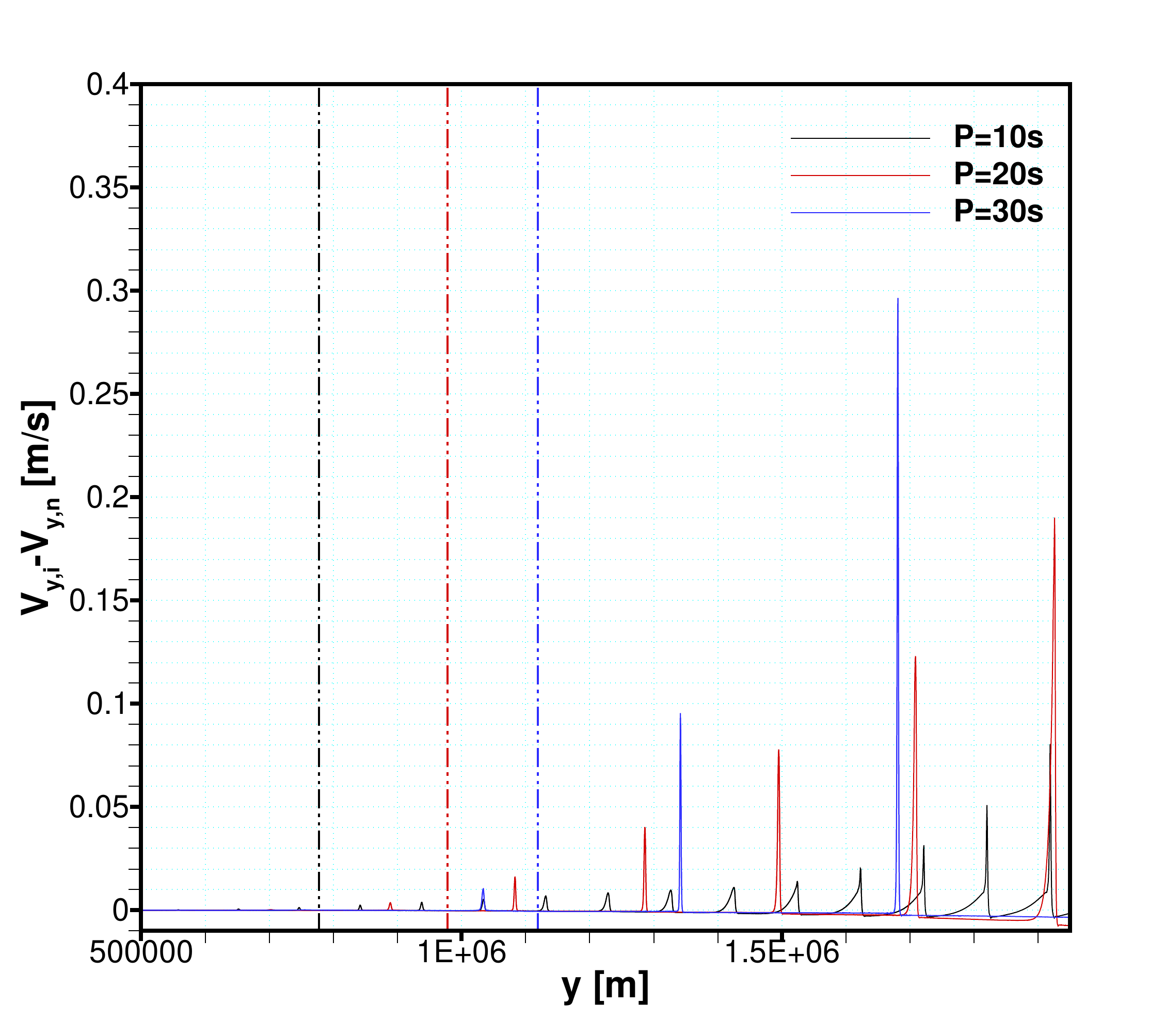}{0.5\textwidth}{(a)}
          \fig{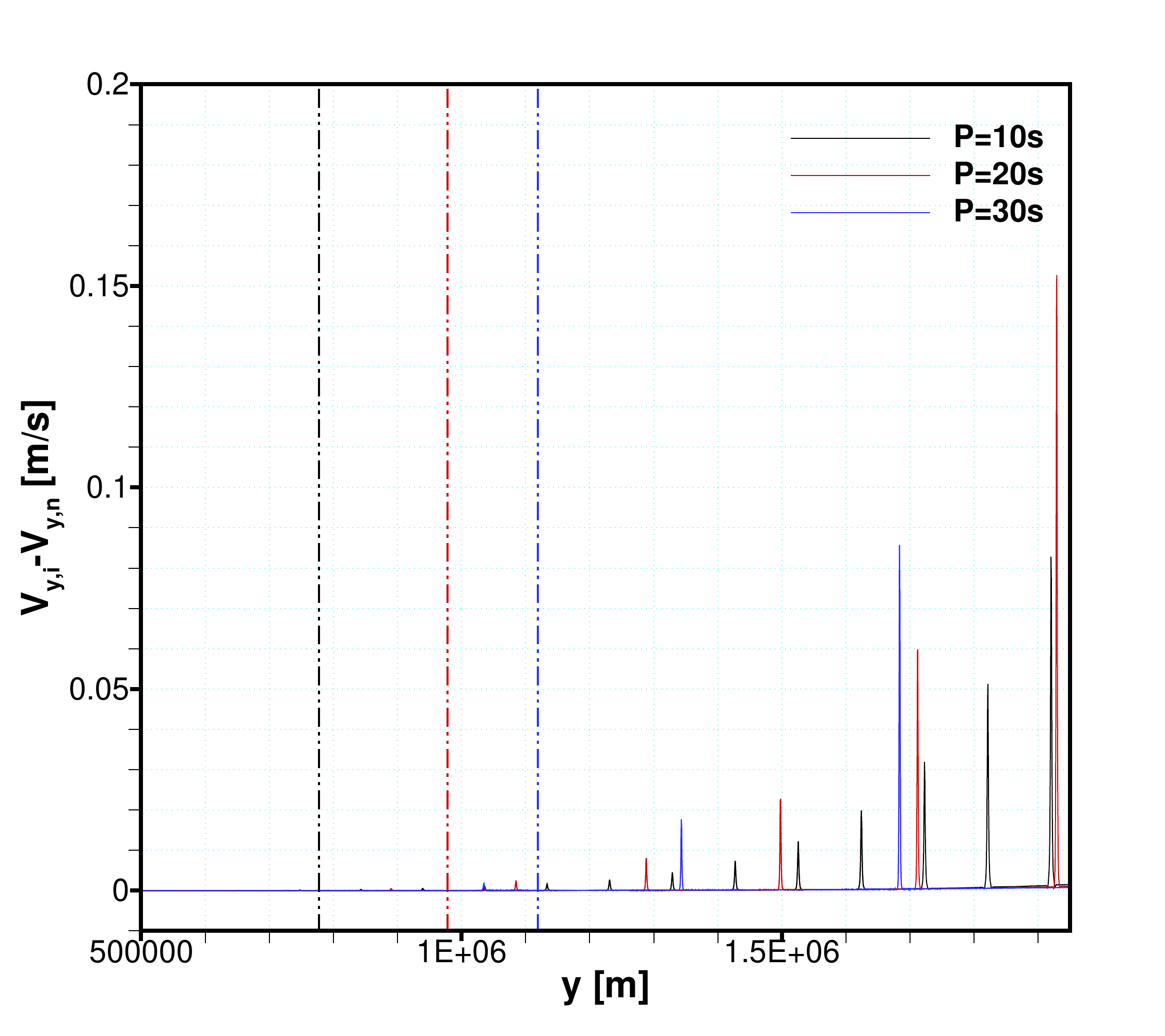}{0.5\textwidth}{(b)}
          }
\caption{Difference between vertical  components of ion and neutral velocities at $t=1200\;$s:
(a) reactive+collisional simulations with initial hydrostatic+chemical equilibrium, (b) collisional simulations with initial hydrostatic+chemical equilibrium. The vertical dashed-dotted lines  indicate the approximate heights at which the strong nonlinear kinetic energy decays start, and each dashed-dotted line corresponds to the solid line of the same colour.
\label{fig:dVReactivedVCollisional}}
\end{figure*}

More information about the decoupling between ions and neutrals is shown in Fig.~\ref{fig:dVReactivedVCollisional}, and similar behaviours can be found in both the reactive+collisional and the collisional simulation results.  In general, imposing longer wave periods leads to more significant difference between the ion and neutral velocities, which is an obvious consequence of the stronger shocks {and indicates the enhanced decoupling between ions and neutrals. However, the difference between the reactive+collisional simulation results and the collisional simulation results is significant. Although the velocity discontinuities (at shocks) are almost the same if the wave period is the same, the velocity differences are significantly larger in the reactive+collisional simulations. Again, we believe this is caused by the momentum exchanges in the ionization and recombination processes, and by the local imbalance resulting from the change of ionization fraction, as mentioned in the last subsection. Apparently, the ionization and recombination processes take place drastically behind the shocks where the temperature is significantly changed.}

\begin{figure*}
\gridline{
          \fig{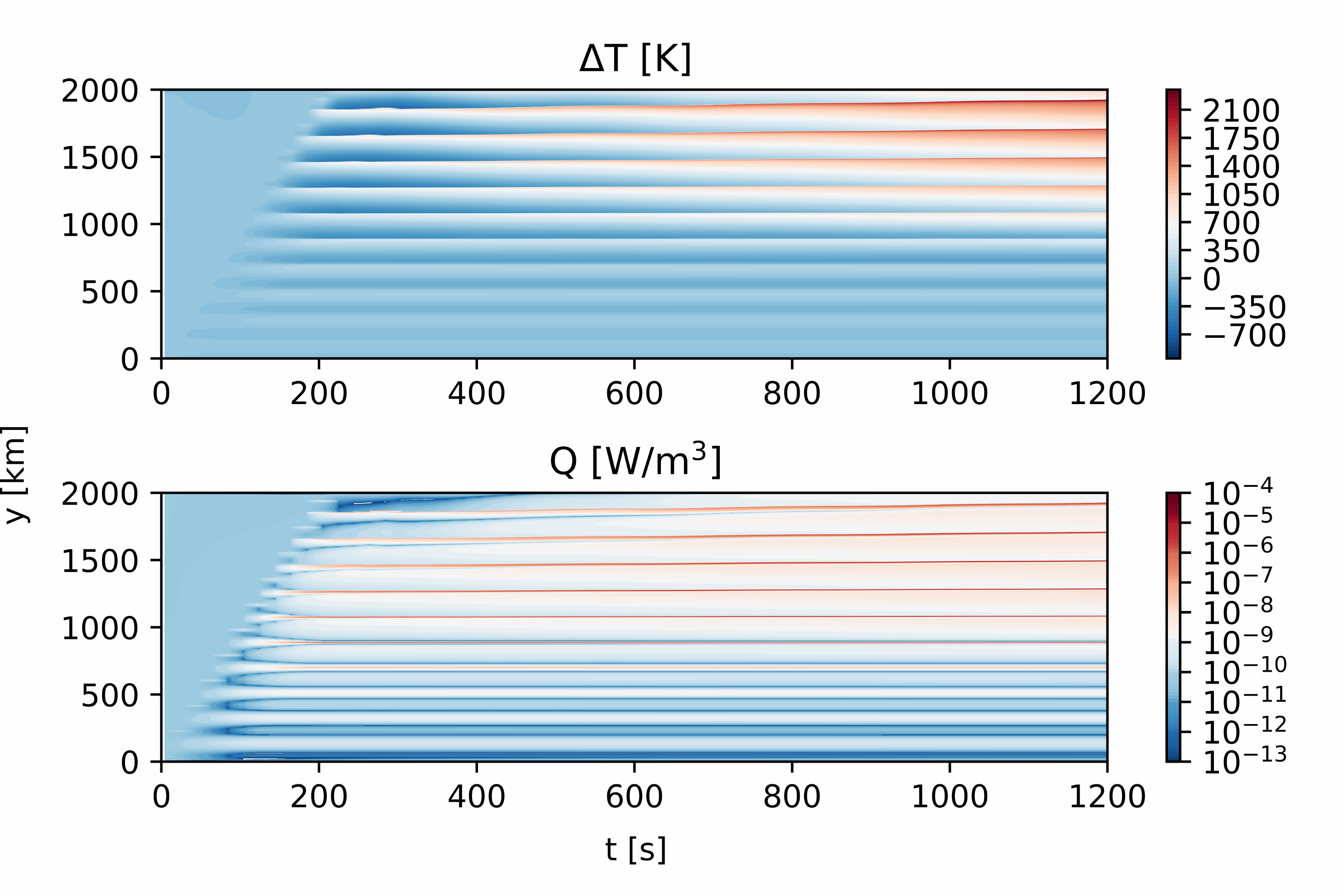}{0.5\textwidth}{(a)}  
          \fig{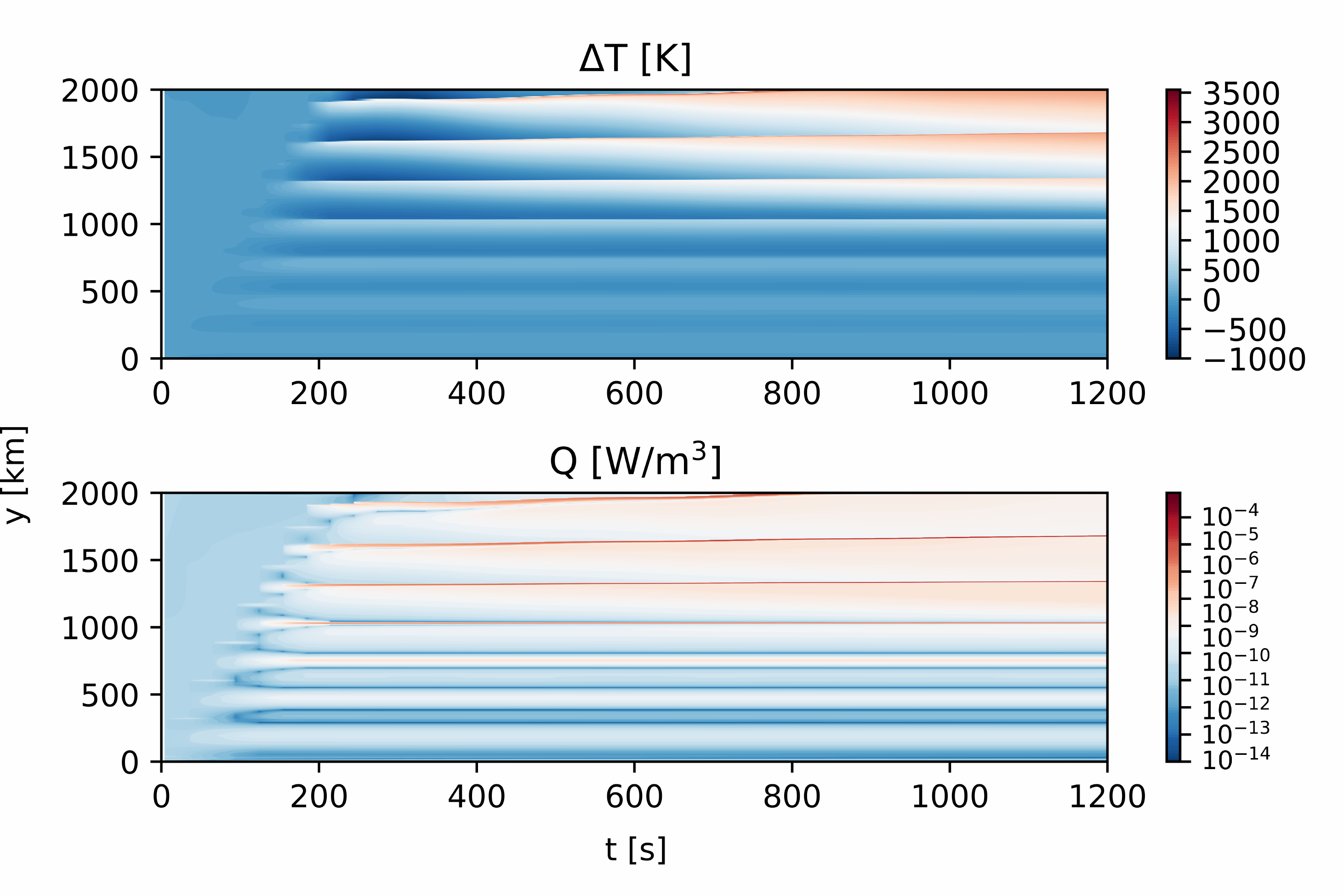}{0.5\textwidth}{(b)} 
          } 
\gridline{ 
           \fig{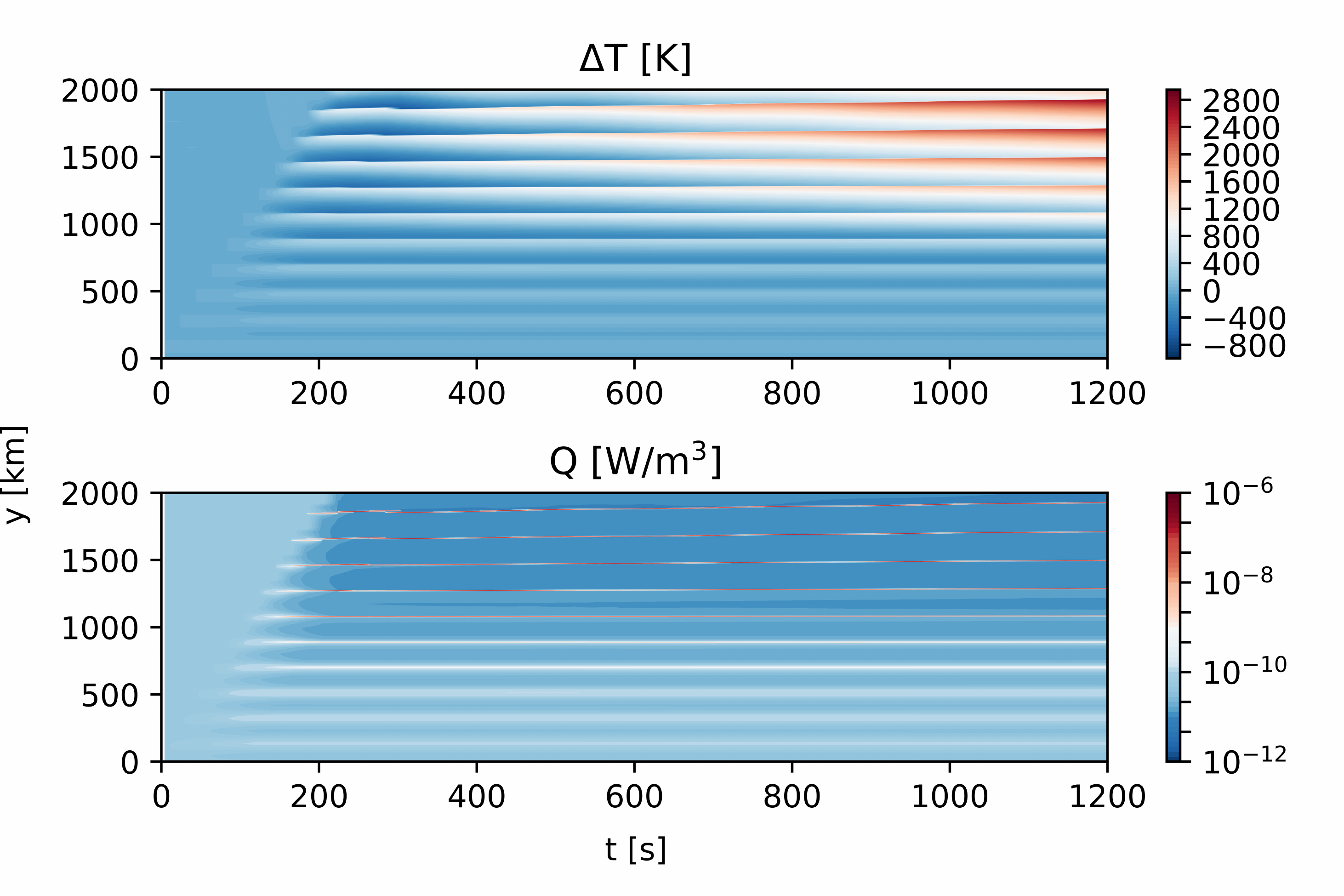}{0.5\textwidth}{(c)} 
           \fig{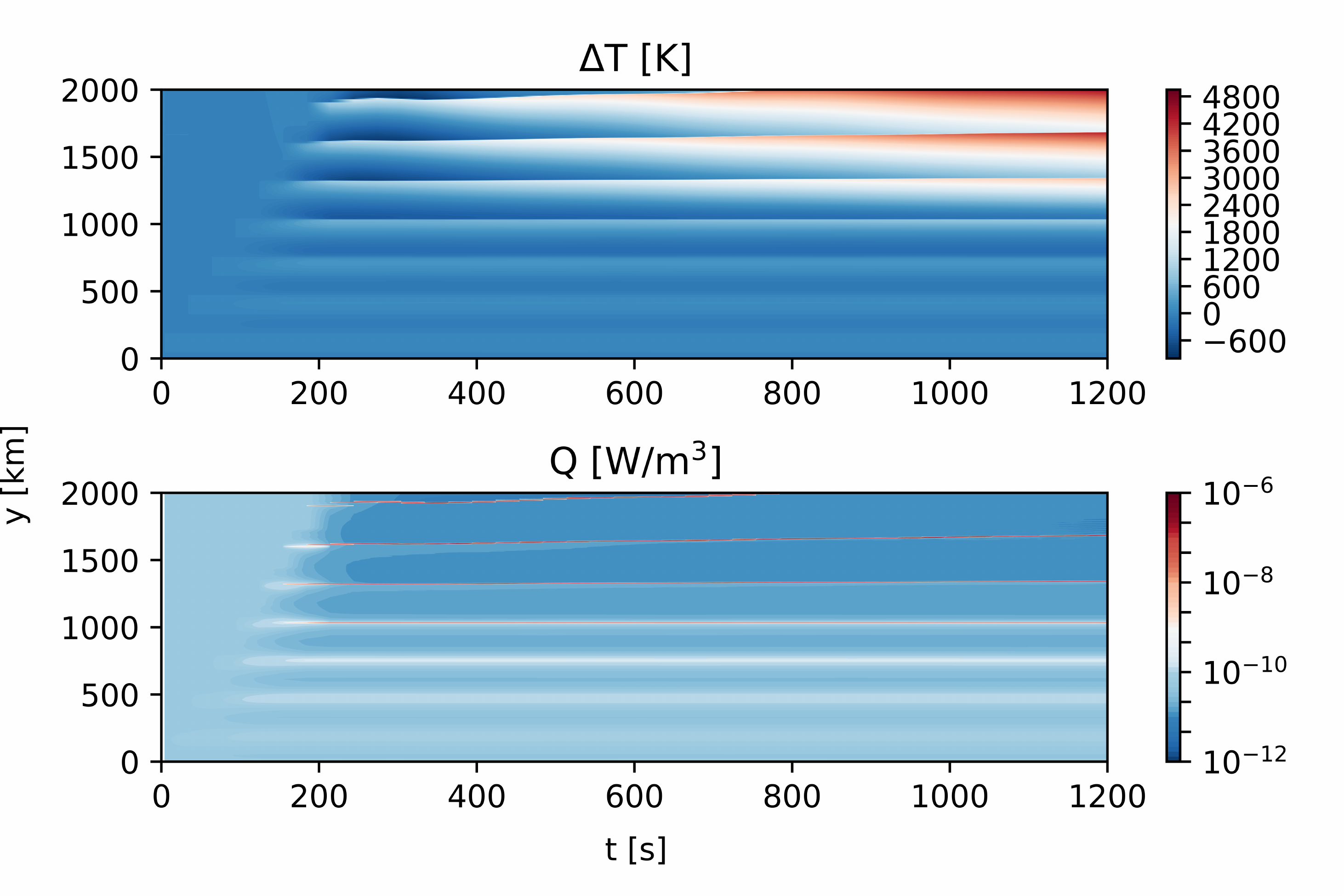}{0.5\textwidth}{(d)}
          }
\caption{{Time-height plots of the net temperature increases and the collisional heating rates: reactive+collisional simulations with initial hydrostatic+chemical equilibrium, and (a) wave period  $P=20\;$s, (b) wave period $P=30\;$s; collisional simulations with initial hydrostatic+chemical equilibrium,  and (c) wave period $P=20\;$s, (d) wave period $P=30\;$s. }
\label{fig:TAPlots_H2}}
\end{figure*}

{In Fig.~\ref{fig:TAPlots_H2}, the time-height plots provide more information about the collisional heating which is a result of the decouping between ions and neutrals discussed in the last paragraph, and the overall temperature increases are also shown. Again, the collisional heating of the reactive+collisional simulations is about two orders of magnitude higher than that of the collisional simulations. Moreover, in the collisional simulations, significant decoupling and collisional heating are both only found at shock fronts. This is not surprising since the minimum collisional frequency ($459\;$s$^{-1}$) is several orders of magnitude higher than the wave frequencies, which means that only at shocks the scale is sufficiently shortened to cause significant decoupling. Whereas, 
since decoupling is also found in smooth regions behind shocks in the reactive+collisional simulations, the corresponding collisional heating is also enhanced, compared with the collisional simulations without the ionization and recombination. However, again, as the ionization process requires a significant amount of energy, the temperature increases in the reactive+collisional simulations are much lower than those of the collisional simulations.}

Then we further discuss in more details about the heating process. As mentioned in the last subsection, the shock heating is likely to be the major reason that causes wave damping and heating, which need to be discussed more quantitatively. Here, we already have the accurate collisional heating rates of all the numerical simulations, by simply calculating the collisional heating source term. Whereas, the shock heating cannot be explicitly and accurately provided as a source term. {Therefore, in order to estimate the shock heating, the energy dissipated by shocks and the heating rates of shocks can be respectively given as} 
\begin{eqnarray}
&\Delta Q\approx T_0\Delta S=T_0 C_{\text{v}} \ln{\frac{T_1/\varrho_1^{\gamma-1}}{T_0/\varrho_0^{\gamma-1}}}, \nonumber \\
&Q_{\text{Shock}}=\frac{\Delta Q }{P}\varrho_0, 
\end{eqnarray}
\noindent {where the subscripts 0 and 1 indicate the quantities ahead of and behind a given shock wave, respectively, $C_{\text{v}}$ is the specific heat at constant volume, and $\Delta S$ is the entropy produced by the shock \citep{Serrin1961}. The (shock) heating rate $Q_{\text{Shock}}$} is a function of the energy dissipated by the shock, the density, and the wave periods.
 More specifically, $T/\varrho^{\gamma-1}$ is directly calculated and then observed based on the numerical results. Because of the limited resolution of capturing shock waves, the observed values are apparently not accurate, but for the estimation they provide enough information in the sense of order of magnitude, as shown in Table~\ref{tab:Heating}.

\begin{deluxetable*}{cccccc}
\tablenum{2}
\tablecaption{{The approximate shock heating rates in regions above $y= 1900\;$km and the approximate maximum collisional heating rates [W/m$^3$]} \label{tab:Heating}}
\tablewidth{0pt}
\tablehead{
  & Shock heating & Collisional heating & Collisional heating\\   
Ionization \& recombination  & No              &   No                 & Yes }
\startdata 
$P=20\;$s   &  $2\times10^{-4}$    & $3\times10^{-7}$  & $3\times10^{-5}$ \\ 
$P=30\;$s   &  $9\times10^{-4}$    & $9\times10^{-7}$  & $2\times10^{-4}$ \\
\enddata
\tablecomments{{The shock heating for wave period $P=10\;$s is not given because the increase of variable $T/\varrho^{\gamma-1}$ could not be numerically resolved. The heating rate for  wave period $P=20\;$s is also under-resolved.}}
\end{deluxetable*}

{It should be noted that the collisional heating rates shown in Table~\ref{tab:Heating} are the maximum peak values of collisional heating, which are found at the shock fronts, and thus the actual collisional heating rates are lower. 
We can see that a shock heating rate could be three orders of magnitude higher than the corresponding collisional heating rate. 
Therefore, according to our estimation, the shock heating is significantly higher than the collisional heating. Although in the reactive+collisional simulations the decoupling between ions and neutrals and the collisional heating are both stronger, the energy dissipated by shocks still dominates the heating process. Therefore, this result supports the explanation of shock heating.} 

\begin{figure*}
\gridline{\fig{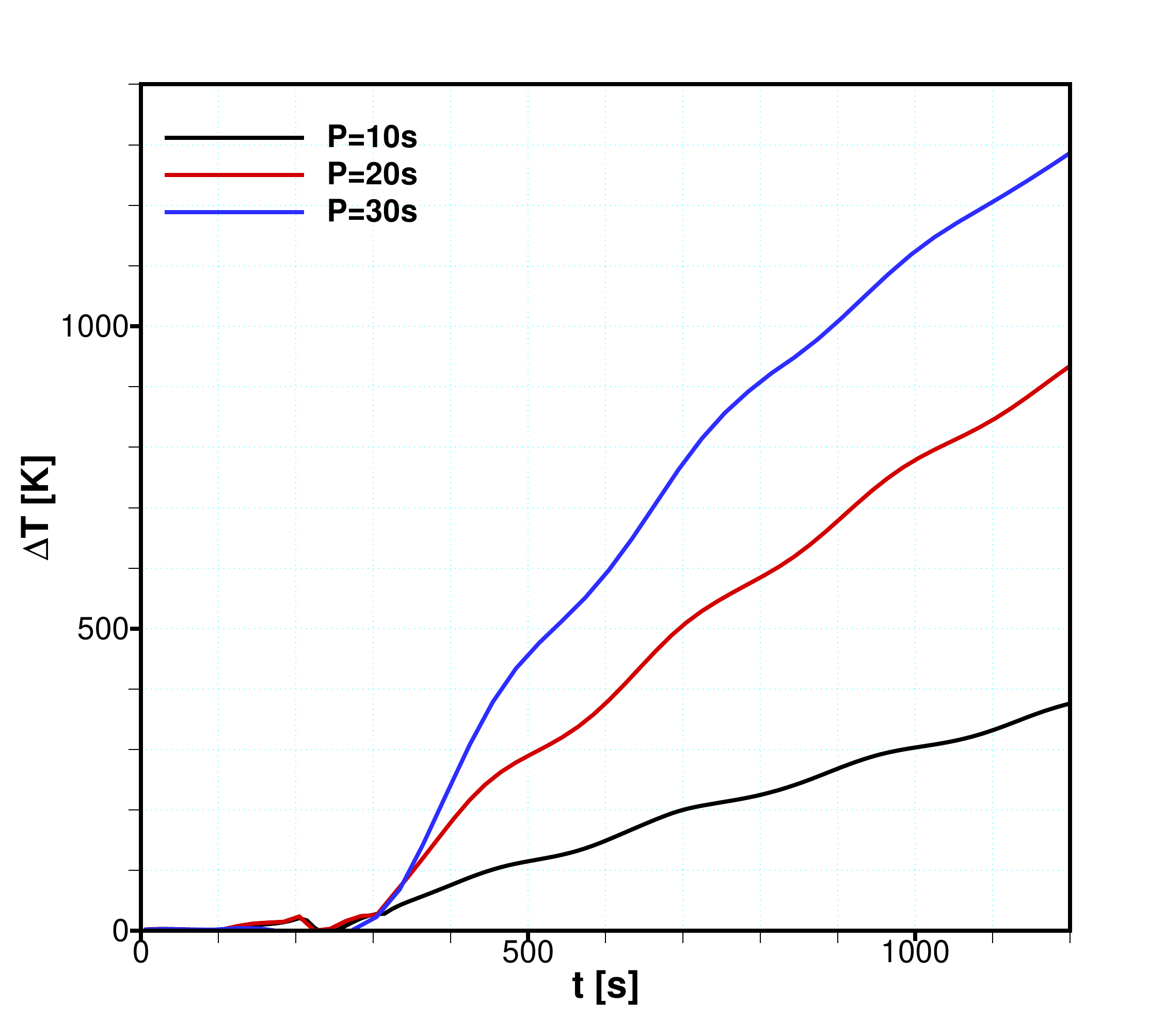}{0.5\textwidth}{(a)}
          \fig{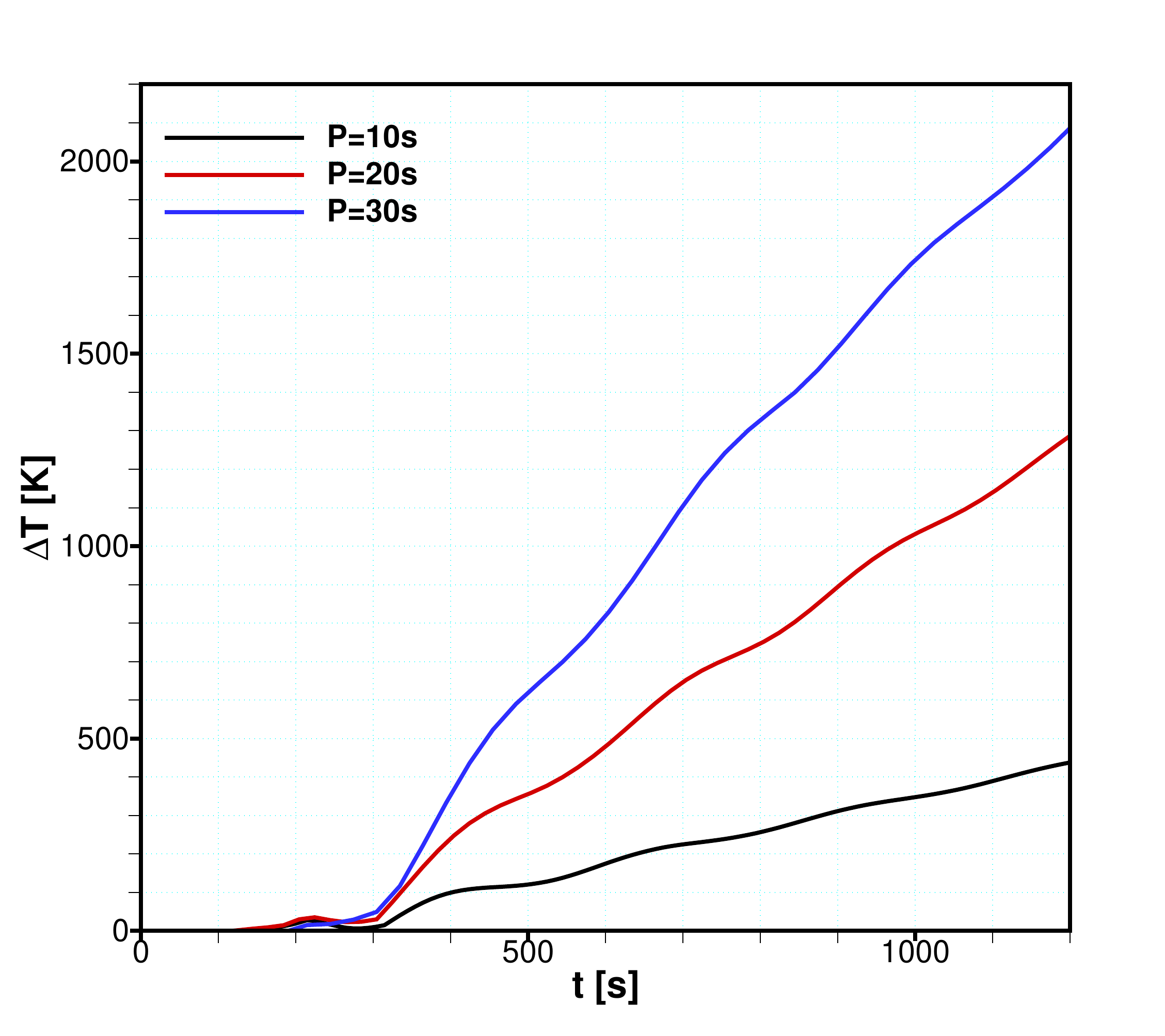}{0.5\textwidth}{(b)}
          }
\caption{{Spatially averaged ion temperature increases ($\Delta T$) over the region of ($1000$ km $ \le y \le 2000\;$km) versus time: (a) reactive+collisional simulations with initial hydrostatic+chemical equilibrium, (b) collisional simulations with initial hydrostatic+chemical equilibrium.}
\label{fig:TimeTemperature}}
\end{figure*}

\begin{figure*}
\gridline{\fig{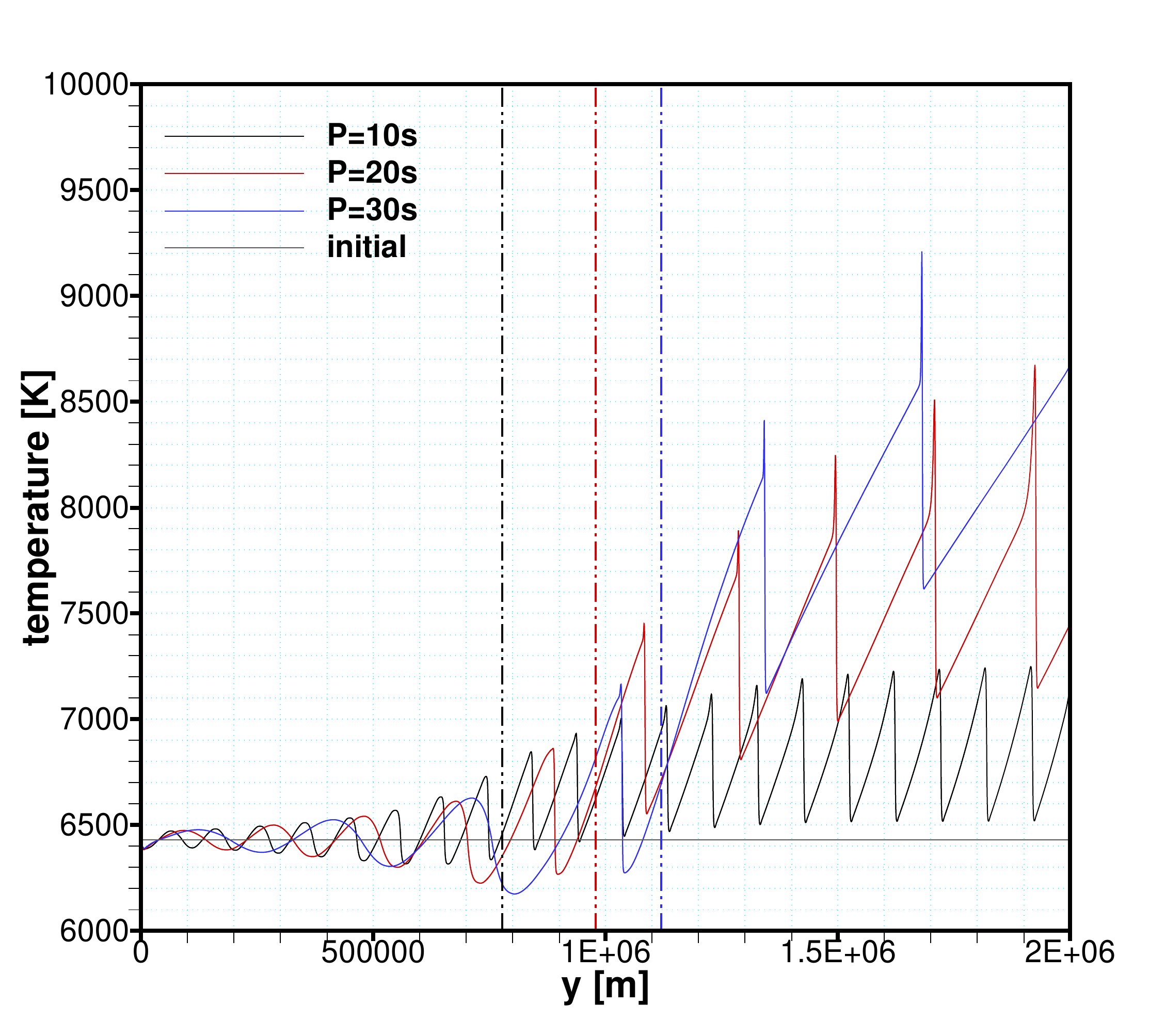}{0.5\textwidth}{(a)}
          \fig{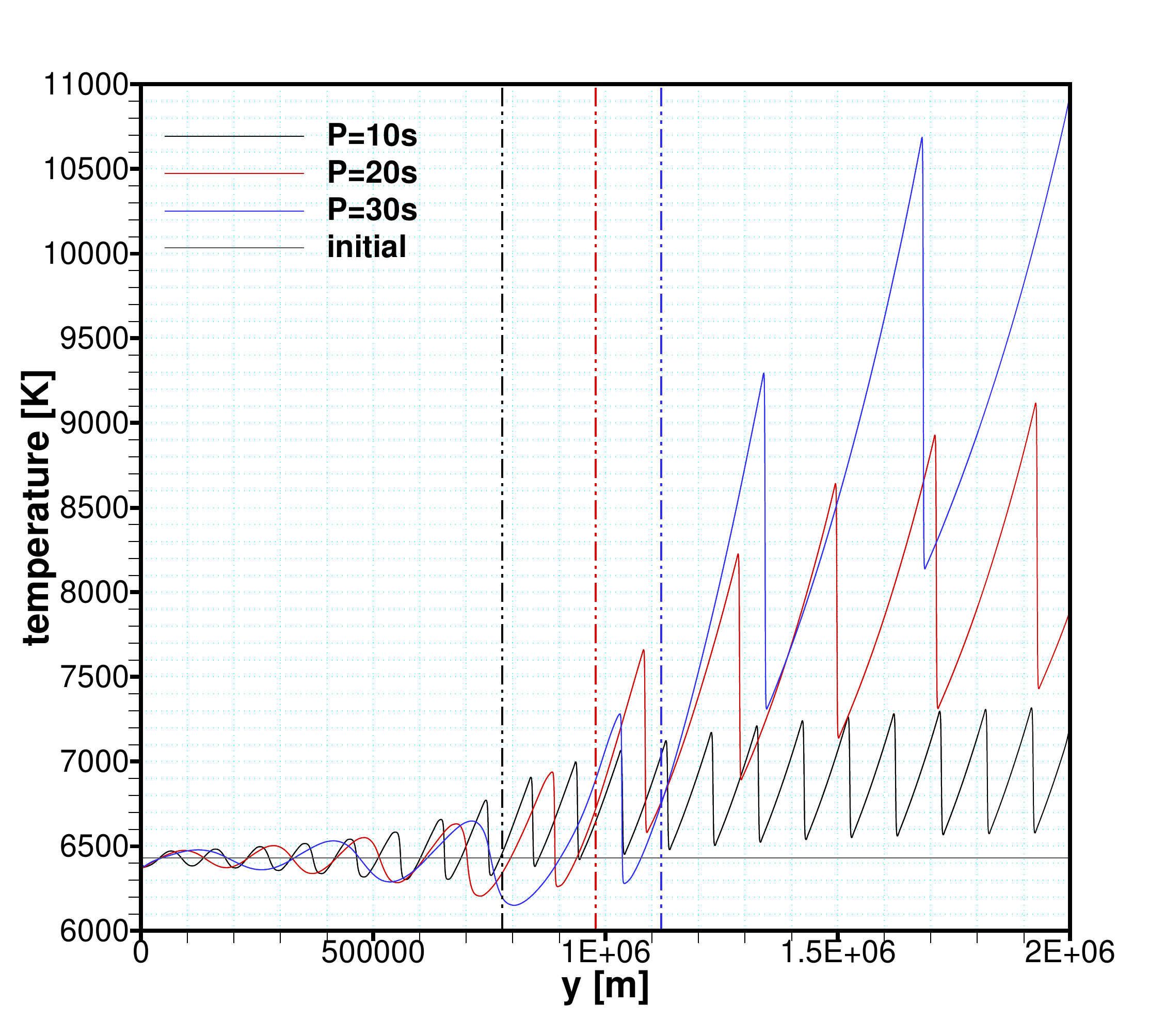}{0.5\textwidth}{(b)}
          }
\caption{Ion temperature profiles of vertical acoustic waves  at $t=1200\;$s: (a) reactive+collisional simulations with initial hydrostatic+chemical equilibrium, (b) collisional simulations with initial hydrostatic+chemical equilibrium. The vertical dashed-dotted lines indicate the approximate heights at which the strong nonlinear kinetic energy decays start, and each dashed-dotted line corresponds to the solid line of the same colour.
\label{fig:TReactiveTCollisional}}
\end{figure*}

{While discussing the difference between simulations with or without the ionization and recombination, it is known that the ionization process may reduce the heating efficiency as it requires a significant amount of energy \citep{Stein1972,Stein1973}. In the present results (Fig.~\ref{fig:TimeTemperature}), we also observe that while including the ionization (and recombination), the temperature increases of numerical results with $10\;$s, $20\;$s, and $30\;$s wave periods are respectively 15$\%$, 28$\%$ and 38$\%$ lower than in the corresponding collisional numerical results. In particular, the maximum transient temperature of the collisional simulation for wave period $P=30\;$s may be close to $11,000\;$K, and is more than $1000\;$K higher than that of the corresponding reactive simulation (Fig.~\ref{fig:TReactiveTCollisional}). Of course, because here we do not consider radiation losses, the temperature keeps increasing, and thus for stronger heating, which is the result of imposing longer wave periods, the ionization process may slow down the heating process more significantly, which can be seen in Fig.~\ref{fig:TimeTemperature}(a). In the reactive simulation with wave period $P=30\;$s, the heating process is clearly slowing down. Whereas, it is worth noting again that the decoupling and the collisional heating are also enhanced while including the ionization and recombination, which were not discussed in previous research.
This enhancement might be more interesting if the wave frequencies are close to the collisional frequency, and/or if a different ionization model is applied.}

\begin{figure*}
\gridline{\fig{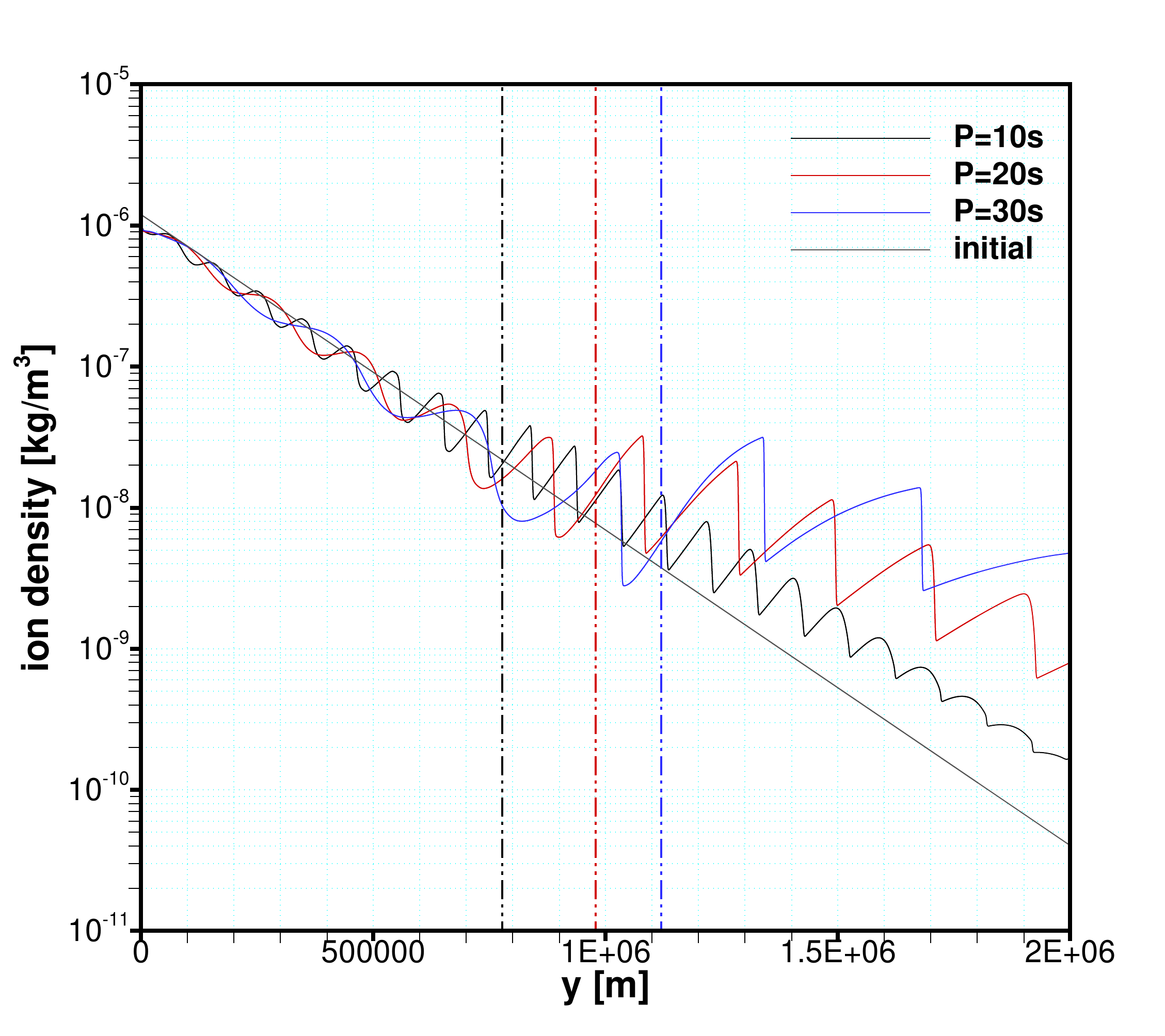}{0.5\textwidth}{(a)}
          \fig{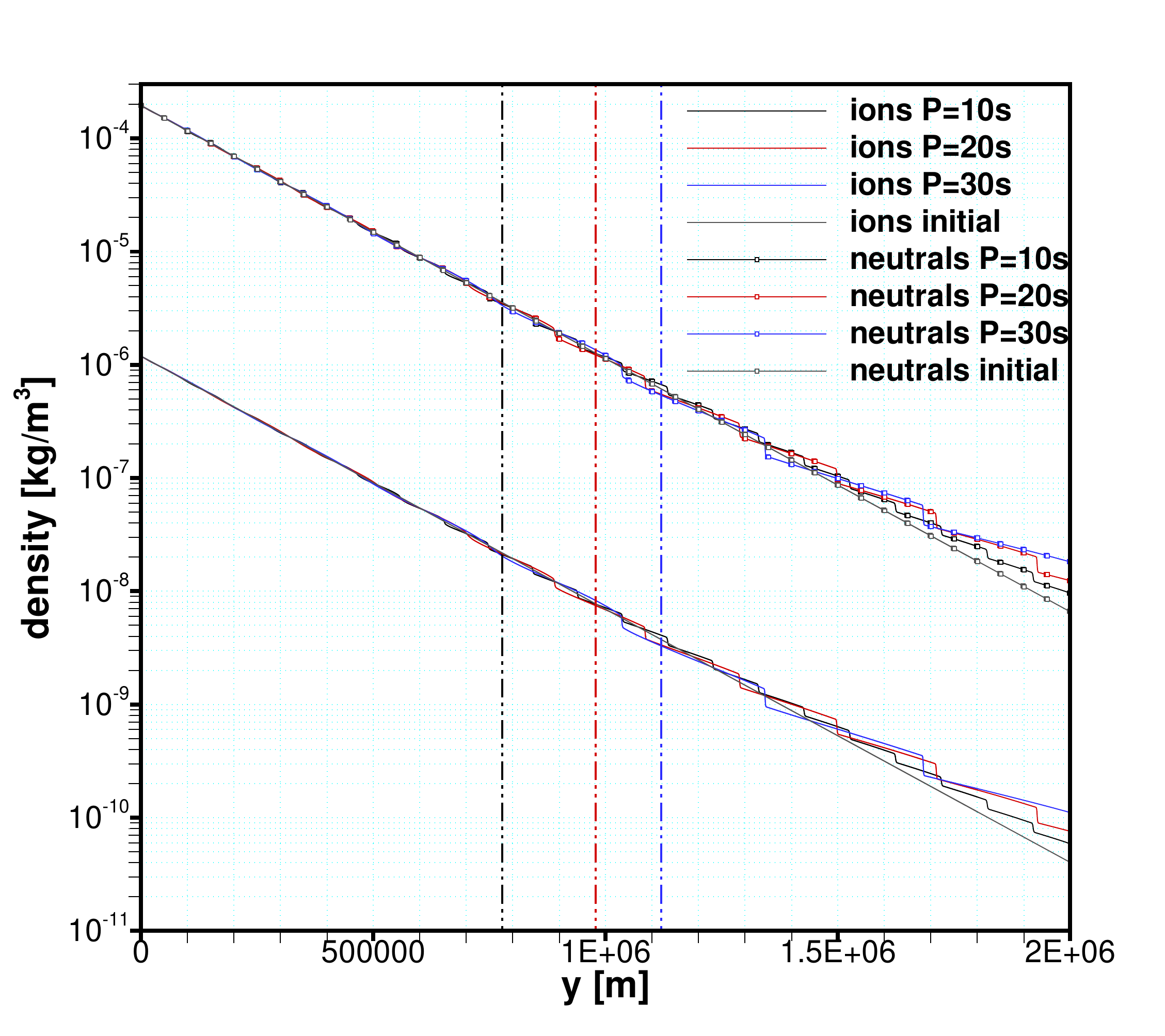}{0.5\textwidth}{(b)}
          }
\gridline{\fig{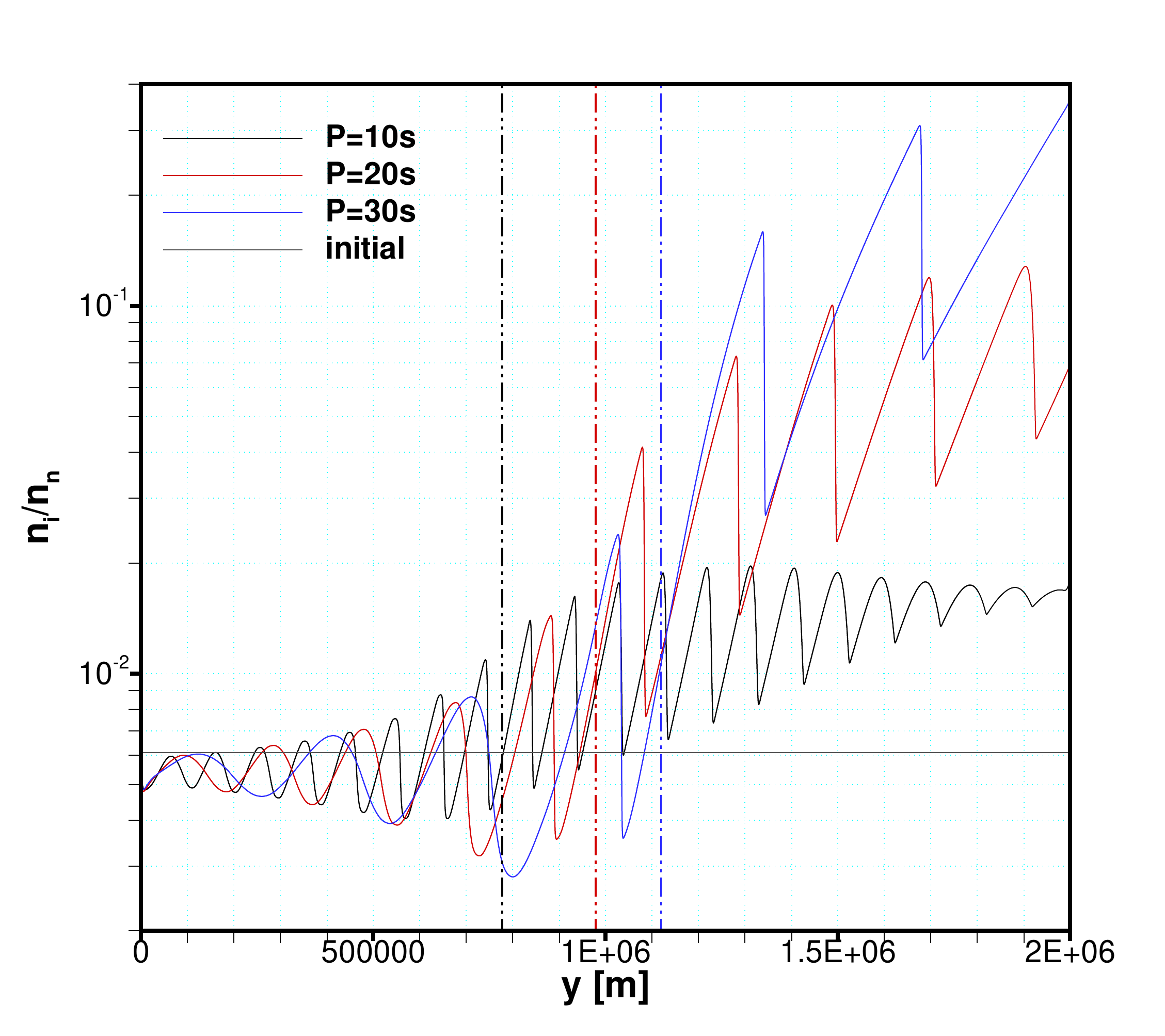}{0.5\textwidth}{(c)}
          }
\caption{Density and ionization fraction profiles of vertical acoustic waves  at $t=1200\;$s: (a) \& (c) reactive+collisional simulations with initial hydrostatic+chemical equilibrium, (b) collisional simulations with initial hydrostatic+chemical equilibrium.  The vertical dashed-dotted lines indicate the approximate heights at which the strong nonlinear kinetic energy decays start, and each dashed-dotted line corresponds to the solid line of the same colour.
\label{fig:RhoReactiveRhoCollisional}}
\end{figure*}

{The results of the ionization process is further shown in Fig.~\ref{fig:RhoReactiveRhoCollisional}. Apparently,} in the collisional simulation results (Fig.~\ref{fig:RhoReactiveRhoCollisional}(b)), the ion density changes due to strong shocks, but the ionization fractions are constant throughout the simulations. Whereas, in the results of the reactive simulations, which include the ionization process, the ion density increases significantly (Fig.~\ref{fig:RhoReactiveRhoCollisional}(a)), and the ionization fractions also increase (Fig.~\ref{fig:RhoReactiveRhoCollisional}(c)). Moreover, the increases of the ionization fractions mostly occur in the right side of the vertical dashed-dotted lines, which indicate the heights at which the strong wave damping starts.  In reality, radiation is another mechanism that cools the chromosphere and thus the ionization fraction cannot be infinitely increased. However, more realistic chromospheric equilibrium is not discussed here.

{Despite the fact that in this work we use a highly idealised model which cannot recover all the properties of the realistic solar atmosphere, we compare the present wave damping effects to the radiative energy losses \citep{Withbroe1977,Sobotka2016}. In order to do so, we first approximately calculate the kinetic energy flux using the formula below:}
\begin{eqnarray}
F(y,t)\approx \frac{1}{2}c_{\text{in}}(y)\left[\varrho_{\text{i}}(y,t)v^2_{\text{i}}(y,t)+\varrho_{\text{n}}(y,t)v^2_{\text{n}}(y,t)\right],
\end{eqnarray}
\noindent {where the sound speed $c_{\text{in}}(y)$ of the initial partially ionized plasma is defined in Eq.~\ref{eq:soundspeed}, and the density and velocity are taken from the quasi-stationary wave trains at $t=1200\;$s. Then we calculate the kinetic energy losses $\Delta F=|F_{y_1}-F_{y_2}|$ between two given altitudes, and the results are shown in Table~\ref{tab:Loss}. The total chromospheric radiative loss \citep{Withbroe1977} and the middle (850 to 1500 km) chromospheric radiative loss \citep{Sobotka2016} of the quiet Sun are provided for comparison.}

\begin{deluxetable*}{ccccc}
\tablenum{3}
\tablecaption{{The radiative energy losses in the chromosphere and the present numerical kinetic energy losses [W/m$^2$].} \label{tab:Loss}}
\tablewidth{0pt}
\tablehead{
  & Total chromospheric losses & Middle chromospheric losses            }
\startdata 
Radiative losses   &  $4\times10^{3}$ \citep{Withbroe1977}    & 3630 \citep{Sobotka2016}   \\ 
$\Delta F$ ($P=10\;$s)   &  $5.6\times10^{3}$    &$5.3\times10^{3}$   \\ 
$\Delta F$ ($P=20\;$s)    &  $5.5\times10^{3}$ &  $4.4\times10^{3}$  \\ 
$\Delta F$ ($P=30\;$s)    &  $5.1\times10^{3}$    & $3.4\times10^{3}$    \\
\enddata 
\tablecomments{{Only the simulations using the H\&C-profile are discussed here.}}
\end{deluxetable*}

{It is found that in the present numerical simulations, the acoustic waves have deposited sufficient energy to compensate the chromospheric energy losses. In particular, although the low-frequency wave ($P=30\;$s) is able to heat the upper region more significantly, the high-frequency wave ($P=10\;$s) deposits more energy at relatively lower altitudes. Of course, we need to note again that the present kinetic energy losses are highly idealised one-dimensional numerical results, without sufficiently taking into account the realistic properties of the lower solar atmosphere.}

\section{Summary and Concluding Remarks}

In this paper, we have performed quasi-1D numerical simulations of acoustic wave propagation in gravitationally stratified and partially
ionized plasmas using a two-fluid plasma(ion)-neutral model, specifically investigating the effects of taking into account the ionization and recombination processes. 
The waves are excited by monochromatic {velocity drivers and then steepen to shocks higher up,} leading to decoupling between ions and neutrals and collisional heating. {While imposing velocity drivers with different wave periods, we have investigated the corresponding heating effects. With a longer wave period, the nonlinear shock wave damping caused by the steepening acoustic wave, may occur at higher altitudes. Before the strong nonlinear damping happens, the kinetic energy of the driven acoustic wave is constant, and a longer wave period leads to the delay of the heating effect, which also means that more energy will be deposited in the higher atmosphere, where the density is much lower and the plasma is easier to be heated. Moreover, a longer wave period reduces the kinetic energy damping rate, and thus the overall heating rate is a result of the interplay between the lower damping rate and the more intensive kinetic energy flux. In general, in the present numerical simulations the kinetic energy is mostly dissipated at shock fronts, and the results are qualitatively similar to those classical acoustic wave and shock wave heating results.}

The major findings and conclusions can be summarised as follows:
 
(I) A gravitationally stratified initial equilibrium model satisfying both hydrostatic and chemical equilibria is provided, and {the ion density in this model significantly differs from typical} hydrostatic equilibrium ion density profiles calculated based on the same reference density {(at the bottom boundary). As a result, in the present numerical simulations, both the collisional heating and the shock wave heating are significantly enhanced compared with the simulations using the hydrostatic equilibrium density profiles. Although the chemical equilibrium assumption might not accurately represent the physics in the realistic lower solar atmosphere, the results still suggest that a more realistic ionization fraction profile may be important for estimating both the shock heating and collisional heating of partially ionized plasmas.}

(II) The ionization and recombination processes are included in modeling acoustic wave propagation in gravitationally stratified and partially ionized plasmas. {As the shock damping dominates the  kinetic energy decays,} these two reactive processes do not directly change the wave damping process, or more precisely, the direct influence is rather small. However, {with the ionization and recombination processes, the decoupling between ions and neutrals and the collisional heating are enhanced. In particular, the collisional heating could be two orders of magnitude higher while involving the ionization and recombination processes. Therefore, the ionization and recombination processes between ions and neutrals are suggested to be taken into account for modeling partially ionized plasmas, and the significance of the processes might be more important if the wave frequencies are close to the collisional frequency. }  

{In conclusion, this work intends to further improve the understanding of wave propagation and damping in partially ionized plasmas by a series of simulations conducted in idealised one-dimensional background fields. Although the present one-dimensional results cannot represent the realistic solar atmosphere as multi-dimensional effects are indispensable for wave propagation, we are able to extend the current understanding of the typical two-fluid numerical modeling in which the collisional interactions are usually investigated. Further research taking into account more realistic ionization models and/or background fields is expected.} 

\acknowledgments

The authors would like to thank the anonymous referee for the comments that greatly helped to improve the manuscript.
These results were obtained in the framework of the projects
C14/19/089  (C1 project Internal Funds KU Leuven), G.0D07.19N  (FWO-Vlaanderen), SIDC Data Exploitation (ESA Prodex-12), and a AFRL/USAF project (AFRL Award No. FA9550-18-1-0093). B.K.'s and K.M.'s work was done within the framework of the projects from the Polish Science Center (NCN) Grant Nos. 2017/25/B/ST9/00506 and 2020/37/B/ST9/00184.
The computational resources and services used in this work were provided by the VSC (Flemish Supercomputer Center),
funded by the Research Foundation - Flanders (FWO) and the Flemish Government - department EWI. 

%

%
%



%
%

\bibliography{ref_MF}
\bibliographystyle{aasjournal}



\end{document}